\documentclass[11pt,letterpaper]{article}
 \pdfoutput=1
 
\usepackage{jcappub}

\usepackage{caption}
\usepackage{floatrow}
\usepackage{feynmp}

\usepackage{subfig}

\usepackage{bm}

\def\be{\begin{eqnarray}}
\def\ee{\end{eqnarray}}

\def\d{\delta_D}

\def\la{\langle}
\def\ra{\rangle}

\title{The Mildly Non-Linear Regime of Structure Formation}
\author{Svetlin Tassev$^{a}$ and Matias Zaldarriaga$^{b}$}
\affiliation{ \sl $^{a}$ Center for Astrophysics, Harvard University, Cambridge, MA 02138, USA\\
 \sl $^{b}$ School of Natural Sciences, Institute for Advanced Study, Olden Lane, Princeton, \\NJ 08540, USA
}

\abstract{We present a simple physically motivated picture for the mildly non-linear regime of structure formation, which captures the effects of the bulk flows. We apply this picture to develop a method to significantly reduce the sample variance in cosmological N-body simulations at the scales relevant to the Baryon Acoustic Oscillations (BAO). The results presented in this paper will allow for a speed-up of an order of magnitude (or more) in the scanning of the cosmological parameter space using N-body simulations for studies which require a good handle of the mildly non-linear regime, such as those targeting the BAO. Using this physical picture we develop a simple formula, which allows for the rapid calculation of the mildly non-linear matter power spectrum to percent level accuracy, and for robust estimation of the BAO scale. 
} 
\notoc
\begin{document}
\maketitle

%\documentclass[11pt,letterpaper]{article}
%\pdfoutput=1
%\usepackage{epstopdf}
%\usepackage{jmy}
%\bibliographystyle{JHEP}

%\usepackage{bm}

\section{Introduction}\label{intro}

The mildly non-linear regime of structure formation has grown in importance recently since the Baryon Acoustic Oscillations (BAO) in the matter power spectrum lie precisely in this regime at low redshift ($z\lesssim 3$), when dark energy comes to dominate. Ever since their detection by the SDSS \cite{2005ApJ...633..560E}, the BAO have become one of the central targets of cosmological investigations. 

\begin{figure}[b!]
\centering
\includegraphics[width=5in]{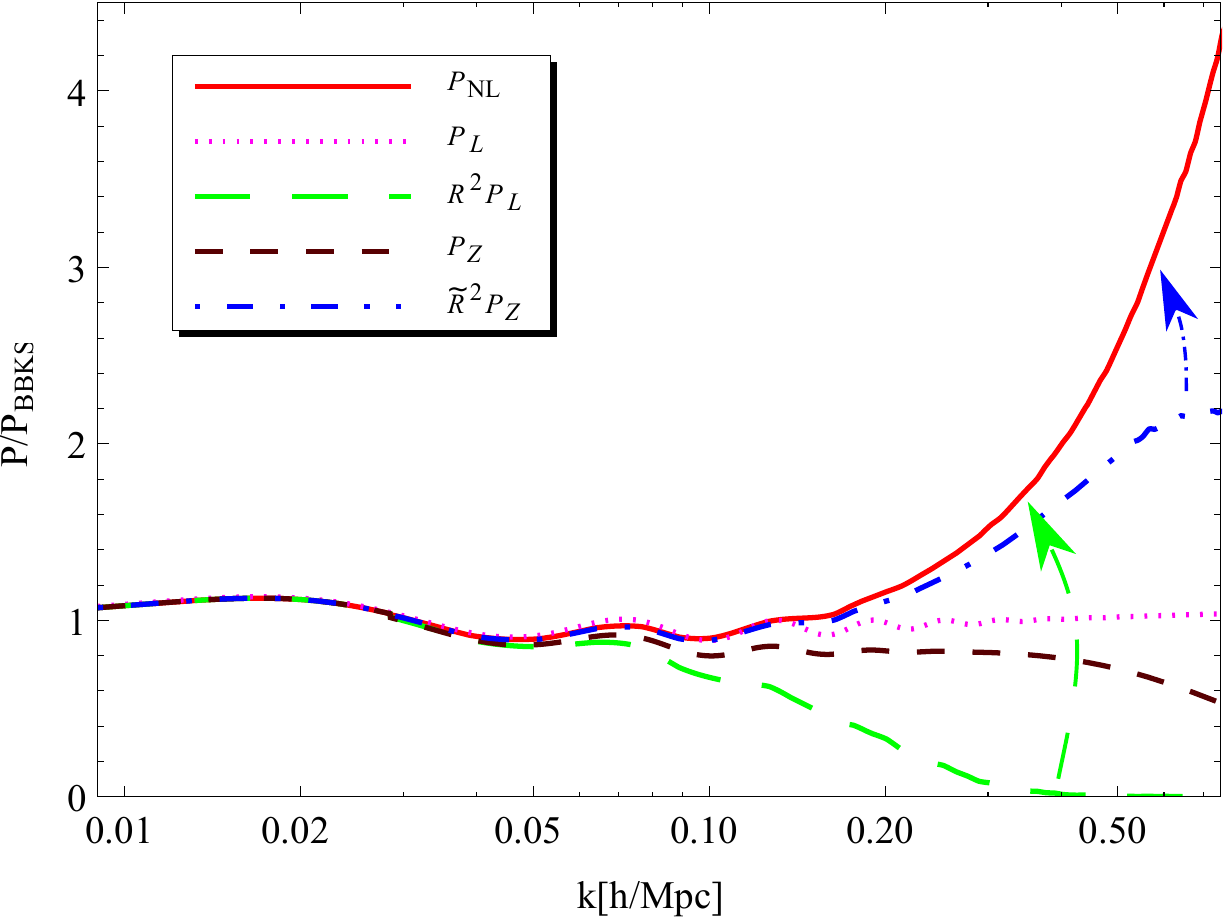}
\caption{\small Shown schematically are various matter power spectra at $z=0$ for $\Lambda$CDM. The power spectra are divided by a smooth BBKS \cite{BBKS} power spectrum with shape parameter $\Gamma=0.15$ in order to highlight the wiggles due to the BAO. The non-linear power spectrum (i.e. the ``exact'' power spectrum obtained from N-body simulations) is given by $P_{NL}$; the linear power spectrum by $P_L$; the part due to the ``memory of the initial conditions'' by $R^2P_L$; the power spectrum calculated in the Zel'dovich approximation by $P_Z$; the power due to the projection of the non-linear density field on the Zel'dovich density field by $\tilde R^2P_Z$. The green (long dashed) arrow  represents the power generated from mode-coupling due to bulk flows, free-streaming and structure formation; while the blue (dot-dashed) arrow represents the power generated by structure formation alone. The current Hubble expansion rate in units of 100\,km/s/Mpc is given by  $h$.} \label{fig:Power}
\end{figure}

The BAO arise  (e.g. \cite{2007ApJ...664..660E}) from the sound waves present in the tightly-coupled photon/baryon fluid in the early universe. After recombination, the radiation pressure can no longer support the baryons, which until then have been driven away from overdensities. Thus, the baryons quickly lose momentum. The resulting comoving sound horizon corresponds to the scale of the BAO, which is $\approx 150\,$Mpc. The BAO in the matter two-point function appear as a peak at that scale with a width of around $20\,$Mpc. That peak translates to a pattern of decaying oscillations in the matter power spectrum with a wavelength of about $0.06\,h/$Mpc (see Figure~\ref{fig:Power}). 

This acoustic signature acts as a standard ruler which may allow us to probe the behavior of dark energy. Any evolution of the equation of state of dark energy may be detectable by a shift in the BAO scale.
Current and future experiments such as BOSS\footnote{http://www.sdss3.org/surveys/boss.php} and WFIRST\footnote{http://wfirst.gsfc.nasa.gov/} are expected to reach 1\% errors in the matter power spectrum at the scales relevant for the BAO; and sub-percent precision in the acoustic scale. 

The acoustic scale is well in the linear regime. Thus, one can expect that the position of the acoustic peak in the 2-point function will be only slightly affected by small-scale physics. However, reaching sub-percent-level accuracy in the acoustic scale at low redshift requires a good handle of the mildly non-linear regime, which (in Fourier space) is probed by the successive peaks and troughs in the power spectrum (Figure~\ref{fig:Power}). One well-recognized effect from the mildly non-linear regime is the suppression of the BAO in the power spectrum due to matter flows, free streaming and non-linear evolution \cite{2007ApJ...664..660E}. These effects introduce a $\sim 10\,$Mpc broadening of the acoustic peak, which in turn degrades the accuracy in the determination of the acoustic scale (in real space). Understanding this degradation can help us construct better peak reconstruction methods, in analogy with \cite{2009PhRvD..79f3523P,2010ApJ...720.1650S}. 

The effects of mode-coupling also introduce a percent-level bias in the location of the acoustic peak at low redshift  \cite{Crocce:2007dt}, exactly where dark energy has an impact. 
 Thus, in order to utilize the coming observations we need to extend our standard cosmological model at least to this mildly non-linear regime (scales of $\sim 10\,$Mpc at $z=0$). 

%Getting a good handle on the transition to the non-linear regime will also have an impact for studying the reionization history of the universe ($z\lesssim 10$) \cite{ref:} which will provide important insights for the first astrophysical sources and late-time structure formation. Going back further, mapping the ``dark ages'' at redshift $6\lesssim z\lesssim 50$ using the 21\,cm transition of neutral hydrogen would allow us to study the evolution of the Intergalactic Medium (IGM) and further test our standard cosmological model \cite{ref:}. 

Going beyond the linear regime is a non-trivial task. Structure formation is a problem without analog, made especially difficult by the free streaming of Cold Dark Matter (CDM) and the long range of the gravitational interaction. Indeed, for arbitrary initial configurations solving the non-linear structure of CDM is a problem much more complicated than turbulence, as the latter can be treated entirely in configuration space as a fluid, using the first several velocity moments of the one-particle distribution function (due to the collisions and short range of the fluid interactions). While a consistent understanding of CDM structure formation in principle requires one to work with the full one-particle distribution function in phase-space.

The reason why structure formation is at all tractable (beyond simulations) at early times is because of the extremely low entropy of its initial conditions (almost homogeneous and isotropic, with almost Gaussian perturbations of the gravitational potential). Moreover, even at late times many interesting phenomena (such as the BAO) are at scales which are either linear or only mildly non-linear. Since particles in Eulerian space have not moved a lot ($\sim 10\,$Mpc on average), at large scales this causes the hierarchy for the velocity moments for the CDM to collapse and one is left with the usual continuity and Euler equations (e.g. \cite{Baumann:2010tm}).

That result can in principle be nullified in the presence of strong backreaction effects. However, recently the effects of the small non-linear scales on the large linear scales have been carefully analyzed analytically in \cite{Baumann:2010tm}, and it was shown that virialized structures completely decouple from large scale modes, while non-virialized small scales introduce only an effective pressure and viscosity to the large-scale equations of motion. Thus, even in the presence of non-linearities, the linear large-scale modes can still be represented as an almost perfect effective fluid.

At the scales relevant for the BAO, Eulerian Standard Perturbation Theory (SPT) (e.g. \cite{jain}) starts breaking down, and thus alternatives must be devised. One way of addressing the problem is through numerical simulations. This has been possible thanks to the exponentially increasing available computational power and the development of successively more efficient algorithms to solve the N-body problem. Thus, conducting cosmological experiments through simulations has proliferated over the last two decades.

However, computational time is still a very important constraint on simulations. Sampling variance requires one to perform numerous simulations of different realizations of the same cosmology; or alternatively simulate large volumes. This prevents an efficient sampling of the cosmological parameter space, and thus, it is still difficult to get a complete handle on the uncertainties that would arise from measurements at the mildly non-linear scales relevant to the BAO (e.g. \cite{Angulo:2007fw}). 
Thus, devising analytical or semi-analytical schemes to study the mildly non-linear regime is of great importance.  Moreover, a good physical picture of what happens in this transitional regime is still needed. That may allow us to construct more robust statistics to study this regime, which are especially needed because what we observe are in fact tracers of the underlying density field, and not the field itself.

\subsection{The Zel'dovich approximation}

A parallel route for studying the mildly non-linear regime is using analytical techniques (e.g. \cite{Crocce:2005xy}). One of the first successful models beyond the fluid approach of Eulerian SPT was the Zel'dovich approximation \cite{zeldovich}, which is easily tractable analytically. In this approximation, the CDM particles are allowed to move on fixed straight trajectories with particle velocities which are drawn from an irrotational stochastic vector field. Particles are allowed to intersect each other's trajectories. After the particle streams start crossing, a rich structure in phase-space develops.

We will argue that the Zel'dovich approximation (ZA) is a crucial stepping stone in understanding the mildly non-linear regime. It is the first order in Lagrangian Perturbation Theory (LPT) (e.g. \cite{buchert,BJCP}) where the expansion parameter is usually taken to be the overdensity, which makes the solutions tractable. In Eulerian space, however, we will see that the ZA is equivalent to certain infinite partial resummations of higher order contributions, which encode some of the effects of the large-scale modes on the mildly non-linear regime.
%Unlike SPT, the ZA does not break down at stream crossing, when an infinity of velocity cumulants are generated. 

\begin{figure}[t!]
\centering
\includegraphics[width=5in]{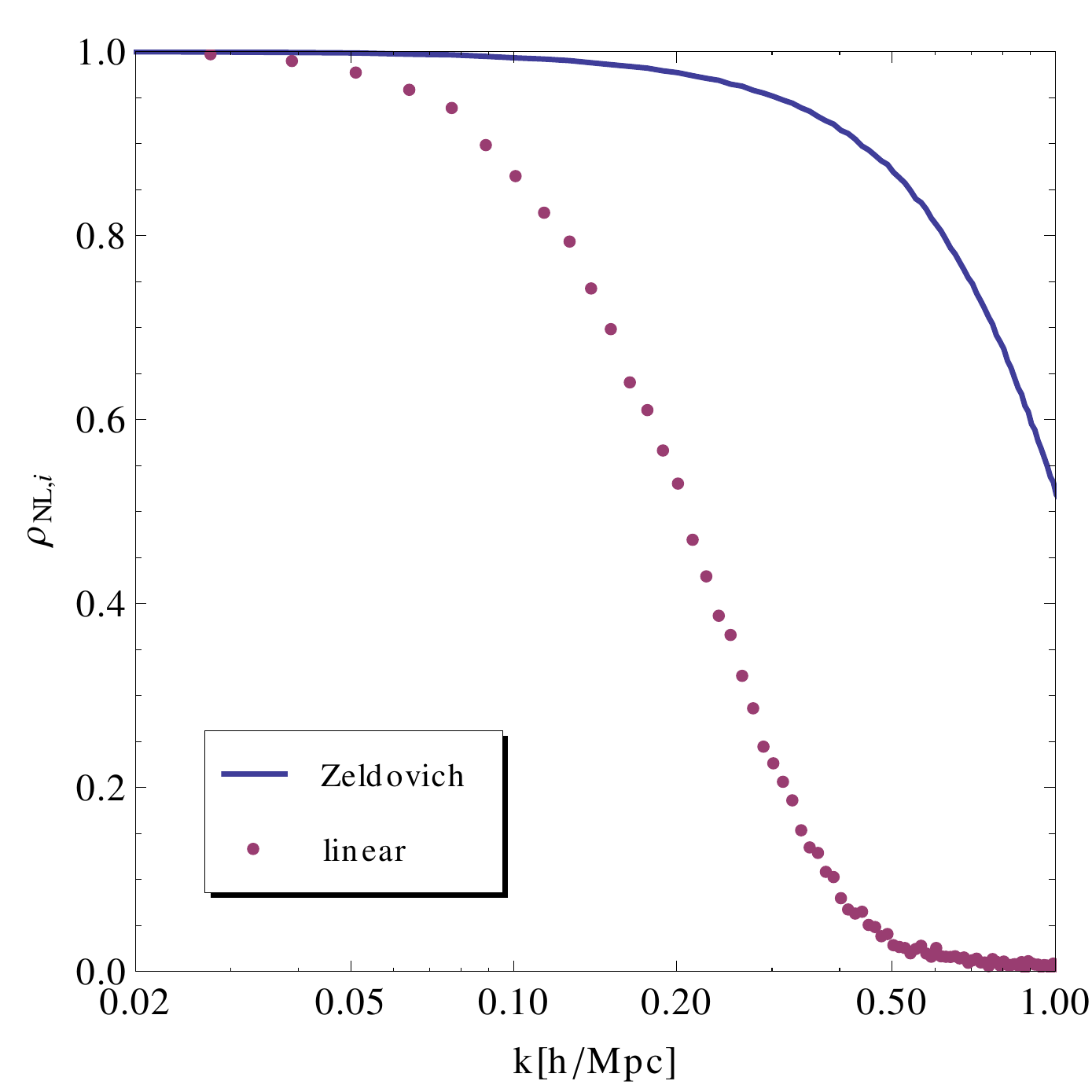}
\caption{\small The cross-correlation coefficient between the non-linear overdensity and the overdensity in linear theory and the ZA as a function of scale at $z=0$. The cross-correlation coefficient, $\rho_{NL,L}$, is very close to the non-linear propagator in RPT, despite their different definitions. Note that the Zel'dovich density field is well-correlated with the non-linear density field well into the mildly non-linear regime.} \label{fig:rho}
\end{figure}

We demonstrate the superiority of the ZA over linear theory\footnote{By ``linear theory'' we refer to the first order SPT.} in Figure~\ref{fig:rho}, where we show the cross-correlation coefficient\footnote{The cross-correlation coefficient between two fields $\delta_a$ and $\delta_b$ is defined as $\rho_{a,b}^2\equiv\la \delta_a\delta_b^*\ra^2/(\la|\delta_a|^2\ra \la|\delta_b|^2\ra)$.} $\rho_{NL,Z}$ between the non-linear overdensity\footnote{By non-linear overdensity and power spectrum we refer to the ``true'' quantities calculated using N-body simulations for example.} and the overdensity calculated in the ZA; as well as the cross-correlation coefficient $\rho_{NL,L}$ between the non-linear overdensity and the overdensity in linear theory. As one can see,  $\rho_{NL,L}$ decays quickly in the mildly non-linear regime; while $\rho_{NL,Z}$ remains close to 1 (see also \cite{1995AA...294..345M}). Given such a high cross-correlation between the non-linear density field and the density field in the ZA, one may ask whether the ZA (and higher order LPT) can be used to correct for the uncertainties associated with cosmic variance at linear and even mildly non-linear scales (corresponding to the BAO) in simulations. This question will be addressed in Section~\ref{sec:correcting}.

Let us now try to understand why linear theory (unlike the ZA) fails to produce a large cross-correlation with the exact result at the mildly non-linear scales, despite the fact that it gives better agreement than the ZA in the power spectrum at these scales (Figure~\ref{fig:Power}). The reason for that will turn out to be the large-scale flows, which are well modeled by the ZA, but not by linear theory.

The effect of the large-scale flows results in corrections in SPT, which are regulated by the parameter $k\sigma_v (<k)$, where $\sigma_v (<k)$ gives the root mean square (rms) particle displacement integrated up to a wavevector $k$  (see also Section~\ref{sec:splitting}).\footnote{This quantity is defined as $(k\sigma_v(<k))^2\equiv k_ik_j\la v_iv_j\ra(<k)=(4\pi/3)k^2\int^k_0P_L(k')dk'$, where $P_L$ is the linear matter power spectrum. This implies that $\sigma_v$ is in fact equal to $(1/\sqrt{3})\times$(rms particle displacements).}
Note that $\sigma_v$ is given by the integral over the velocity power spectrum and is proportional to the velocity dispersion, which is why for brevity we will usually refer to $\sigma_v$ as the velocity dispersion. 
The corrections from the largest-scale flows ($k\to 0$) due to $\sigma_v(<k)$ cancel at each order in SPT \cite{1996ApJ...456...43J}. This must be the case because uniform motions translate coherently the density field, which has no observable effect when equal-time statistics are considered. However, this is no longer true when one considers statistics at different times.

\begin{figure}[t!]
\centering
\includegraphics[width=5in]{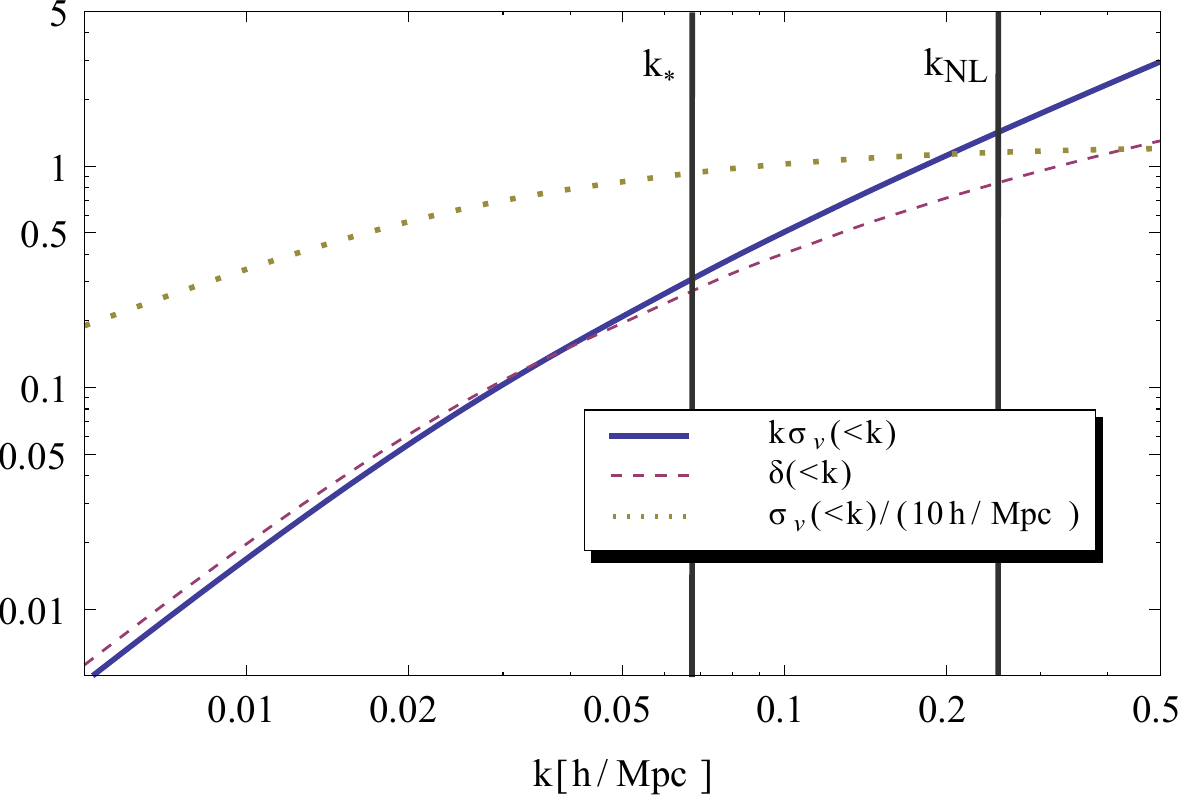}
\caption{\small We show the overdensity $\delta(<k)$ and rms particle displacements $\sigma_v(<k)$ integrated up to $k$, calculated in linear theory at $z=0$. The quantity $k\sigma_v(<k)$  is due to the bulk flows and it regulates the destruction of the cross-correlation between the linear and non-linear density fields. Note that $k\sigma_v(<k)$ is larger than $\delta(<k)$ in the mildly non-linear regime. The peak of the velocity power spectrum is at $k_*$ which is well below the non-linear scale $k_{NL}$ when the linear density power per logarithmic $k$-interval reaches 1. Therefore $\sigma_v$ is roughly constant in the mildly non-linear regime. In the plot we marked  with vertical lines the positions of $k_*$ and $k_{NL}$ for reference. See the text for further discussion. } \label{fig:deltas}
\end{figure}

Linear theory predicts a density field which is simply an overall rescaling of the initial density field. Therefore, $\rho_{NL,L}$ also equals the cross-correlation between the initial and final density fields. This is a statistics at different times and so it is affected by the bulk flows. 
The velocity power spectrum peaks at rather large scales (at $k_*\approx 0.07\, h/$Mpc; see Figure~\ref{fig:deltas}), therefore it is precisely large-scale motions that dominate.
So, in the mildly non-linear regime the density field at a later time can be crudely thought of as a translated version of the initial density field, where the translation is a random variable. Such a uniform random shift destroys the cross-correlation between the initial and final density fields. This explains why $\rho_{NL,L}$ decays so quickly in the mildly non-linear regime.

In Figure~\ref{fig:deltas}  we show the magnitude of $k\sigma_v (<k)$ and the overdensity, $\delta(<k)$, integrated up to $k$ (the latter is defined through eq.~(\ref{Pnlshort})). Since the velocity power spectrum peaks at large scales, the velocity dispersion $\sigma_v (<k)$ reaches a constant well below the non-linear scale, $k_{NL}$, (see Figure~\ref{fig:deltas}). Thus, even though $k\sigma_v (<k)$ grows large in the mildly non-linear regime, its effects should be accounted for by the ZA, which exactly captures the effects of the convective derivative in the Euler equation. 
%Therefore, the effect of the bulk flows is captured by LPT, and thus the ordering parameter in LPT in Eulerian space is given by $\delta(<k)$. 
In the mildly non-linear regime  $k\sigma_v (<k)$ is larger than $\delta(<k)$. Therefore, the equation of motion for the particles in the ZA captures significant information that is encoded beyond linear theory. As we will argue, this can explain the large cross-correlation between the Zel'dovich density field and the true density field.

%\footnote{The corrections in Eulerian space in the mildly non-linear regime are larger for SPT (corrections regulated by both $k\sigma_v(<k)$ and $\delta(<k)$) than for LPT (corrections regulated only by $\delta(<k)$). Therefore, the corrections which are resummed by LPT, but are treated perturbatively in SPT, are larger than the effects which are treated perturbatively in both SPT and LPT (corrections regulated by $\delta(<k)$).} 

%Therefore, one can conjecture that the dynamics in the mildly non-linear regime are better captured under the ZA than linear theory, despite the fact that linear theory gives better agreement in the power spectrum at these scales (Figure~\ref{fig:Power}). 

%Indeed, at the level of the matter power spectrum the first order in the SPT (which for short we will simply call ``linear theory'' from now on) provides good agreement (to 1\%) with the non-linear power spectrum, $P_{NL}$, down to scales of $\sim 100$\,Mpc at $z=0$, while the ZA gives the same level of agreements down to twice smaller scales. This is compensated by the fact that at smaller scales the ZA quickly diverges from the $P_{NL}$ (to the level of $\sim$5\%), while linear theory remains relatively accurate down to $\sim 10\,$Mpc scales (see Fig.~\ref{fig:Power}). 

\subsection{Beyond Standard Perturbation Theory}

Recently various analytical tricks have been put forward to remedy the convergence properties in the weakly nonlinear regime of SPT \cite{jain}. These include renormalized perturbation theory RPT \cite{Crocce:2005xy}, the path-integral approach \cite{Valageas:2006bi}, and the renormalization group flow \cite{Matarrese:2007aj}. Most of these expansion schemes work in the single-stream (or fluid) approximation. Thus, their results are applicable before shell-crossing, i.e. long before virialization can occur. They work  only with the first two velocity moments of the Vlasov equation which reduce to the usual continuity and Euler equations. Since the higher moments of the one-particle distribution function are artificially discarded, one closes the system by introducing an equation of state, or equivalently a sound speed, which is set to zero for the CDM (along with any anisotropic stress). The effects of the pressureless perfect fluid approximation on the power spectrum at the mildly non-linear scales relevant to the BAO at $z=0$ have been estimated to be around 1\% or more (e.g. \cite{Afshordi:2006ch,Pueblas:2008uv,McDonald:2009hs}), which may not be sufficient to satisfy the requirements imposed by observations.\footnote{The effect of these corrections on the acoustic peak has not yet been evaluated, especially because all relevant calculations of the effect have been done using severe approximations.}

Nevertheless, some of these expansion schemes have gained prominence and a discussion of their results is warranted. We will focus on two representative methods, one of them derived starting from the fluid approximation, while the other starting from LPT. Those are the renormalized perturbation theory (RPT) \cite{Crocce:2005xy} and the Lagrangian Perturbation Theory version of  \cite{matsubara}.\footnote{To distinguish the original LPT from its approximated version offered by \cite{matsubara}, we will refer to the latter as MLPT.} In these approaches (and most other recently proposed methods), one writes the non-linear power spectrum, $P_{NL}$, as a sum of two pieces $P_{NL}=R^2P_L +P_{MC}$, where $P_L$ is the power spectrum from linear theory, $R$ is the response function (or ``propagator'') which gives the memory of the initial conditions, and $P_{MC}$ gives the power generated from mode-coupling. This split has been extensively used recently, and has been applied in the reconstruction of the acoustic peak \cite{2009PhRvD..79f3523P,2010ApJ...720.1650S}, and so deserves special attention. 

In the split above, the response function is simply defined as the projection of the non-linear density field on the linear density field, and is therefore proportional to the cross-correlation between the initial density field (which is in turn proportional to the linear density field) and the density field at the moment in time of interest. However, as we already discussed that cross-correlation decays quickly in the mildly non-linear regime ($\rho_{NL,L}$ in Figure~\ref{fig:rho}). Thus, $R$ must decay quickly at small scales as well.\footnote{At high $k$, $R^2$ is well approximated by $\exp(-k^2\sigma_v^2)$ \cite{Crocce:2005xz}, where $\sigma_v\sim 9\,$Mpc at $z=0$ is $(1/\sqrt{3})$ times the rms displacement of the particles in linear theory.} This can be clearly seen in Figure~\ref{fig:Power} where we plot $R^2P_L$.

Written that way, the mode-coupling power, $P_{MC}$, has three sources: bulk flows, shell crossing and structure formation, where (for brevity) we define the latter to refer to all effects not captured by the ZA (see below). So, $P_{MC}$ must compensate for the difference in power denoted by the green (long dashed) arrow in Figure~\ref{fig:Power}.

Let us try to understand the above split when written for the density field calculated in the ZA. The power spectrum in the ZA, $P_Z$, can be similarly split into $P_Z=R_Z^2P_L+P_{MC}^Z$, where a $Z$ indicates that we are working in the ZA; and $R_Z$ is approximately equal to $R$. The ZA leads to a similarly decaying density response function as for the non-linear power spectrum. However, the decay seen in the term $R_Z^2P_L$ is completely absent in the total Zel'dovich power spectrum, $P_Z$ (see Figure~\ref{fig:Power}), being compensated by the mode-coupling power $P_{MC}^Z$ which is due to the effects of the large-scale flows and free streaming. This result can be understood in light of our discussion of the effects of the bulk flows: Although the memory of the initial conditions for the density at small scales is indeed lost under the random large-scale flows, the small-scale density correlations evaluated at the same moment in time are in fact preserved. Indeed, in the ZA one can show (Section~\ref{sec:splitting}) that the split above is due to the ability to choose different ordering parameters, over which to expand $P_Z$. The expansion parameter resulting in the split of $P_{Z}$ above is in fact not small, since it is $\sim\sigma_v^2k^2\sim 1$ at the scales relevant for the BAO, where  $\sigma_v\approx 9\,$Mpc at $z=0$ is related to the rms displacement of the particles in linear theory. So, the splitting of $P_Z$ into $R_Z^2P_L+P^Z_{MC}$ is not perturbative for the mildly non-linear regime, and this argument can analogously be extended to the splitting of $P_{NL}$ in RPT (see also Section~\ref{sec:splitting}).
To put it in another way, RPT and MLPT split the effect of the bulk flows, parametrized by $k\sigma_v(<k)$, between the pieces $R^2P_L$ and $P_{MC}$. However, as discussed above, the effects of the large-scale flows must cancel for equal-time statistics -- something which is no longer true order by order in RPT and MLPT, although it must hold non-perturbatively.

Thus, by using the density response function, both RPT and MLPT treat the cancellation of the effects of the bulk flows as something to be recovered only at higher orders. 
This problem can be circumvented by realizing that information about the large-scale flows (as well as some of the effect of shell crossing\footnote{The equation of motion corresponding to LPT treats the density in the Poisson equation in the single-stream approximation. However, the resulting particle trajectories are still allowed to intersect, resulting in a rich structure in phase-space. It is in this sense that we say that LPT, and the ZA in particular, capture some of the shell-crossing information but not all of it.}) can be easily restored by working in the ZA.

In this paper we propose a different, physically motivated way to split the effects of mode-coupling, which will allow us to construct a simple model for the mildly non-linear regime. 
If one is able to collect the mode-coupling power due to the bulk flows with the power due to the memory of the initial conditions, $R^2P_L$, one may be able to write $P_{NL}=\tilde R^2P_Z+\tilde P_{MC}$, with a new $\tilde R$ and new $\tilde P_{MC}$ which are due to structure formation (by our definition).  Here we defined $\tilde R$ as the projection of the non-linear field on the Zel'dovich density field. Since the ZA captures the effects of the bulk flows\footnote{As shown by \cite{1996ApJS..105...37S}, the ZA satisfies Galilean invariance for the continuity equation and the divergence of the Euler equation.}, and $\tilde R$ is proportional to $\rho_{NL,Z}$ (the factor of proportionality depending on equal-time statistics), we can conclude that all three quantities, $\tilde R$, $P_Z$ and $\tilde P_{MC}$, are independent of the large-scale flows. Indeed, the Zel'dovich density field is well-correlated with the non-linear density field well into the mildly non-linear regime  (see Figure~\ref{fig:rho}), and therefore, the two are simple $k$-dependent rescalings of each other. Thus, in Figure~\ref{fig:Power} one can clearly see that $\tilde R^2P_Z$ follows $P_{NL}$ closely even for $k\sim k_{NL}$. The new $\tilde P_{MC}$ must be such as to compensate for the difference in power indicated by the blue (dot-dashed) arrow in Figure~\ref{fig:Power}. So, we find that $\tilde P_{MC}$ is nearly irrelevant for the BAO (Figure~\ref{fig:Power}). Judging from the smaller mode-coupling power that needs to be obtained, calculating $P_{NL}$ in the mildly non-linear regime using the split we proposed, may be easier than the route taken by RPT.

%==========================================
%MECHKAMECHKAMECHKAMECHKAMECHKAMECHKAMECHKAMECHKAMECHKAMECHKAMECHKAMECHKAMECHKAMECHKAMECHKA

In Section~\ref{sec:splitting} we further discuss the proposed split of $P_{NL}$. In Section~\ref{sec:correcting} we demonstrate how the proposed split can be applied in correcting N-body simulations for sample variance in the linear and mildly non-linear regimes. We show that the results presented in this paper will allow for a speed-up of an order of magnitude (or more) in the scanning of the cosmological parameter space for studies which require a good handle of the mildly non-linear regime, such as those targeting the BAO. In Section~\ref{whyWork} we explain why the method we propose for eliminating the sample variance from simulations works so well. Then in Section~\ref{sec:model} we construct a simple physically motivated model for the mildly-nonlinear regime, which captures the effects of bulk flows\footnote{In the sense that bulk flows are irrelevant for equal time-statistics order by order in the proposed split of $P_{NL}$.}; and which models the matter power spectrum to 1\% accuracy for these scales and gives robust estimates for the acoustic scale. We then summarize our results in Section~\ref{sec:summary}. In Appendix~\ref{appA} we provide some more details on the derivation of our model.

\section{Splitting the non-linear power spectrum}\label{sec:splitting}

As we discussed in Section~\ref{intro}, many recently proposed expansion schemes targeting the mildly non-linear regime, such as the RPT and MLPT expand the non-linear power spectrum, $P_{NL}$, as a sum of two terms -- a term proportional to the linear power spectrum, $P_L$, and a remainder, called the mode-coupling term, $P_{MC}$:
\be\label{split}
P_{NL}(k)=R^2(k,\eta) P_L(k,\eta)+P_{MC}(k,\eta)\ ,
\ee
where $\eta$ is conformal time. The response function, $R$ is given by the projection of the non-linear density field, $\delta_{NL}$, on the linear density field, $\delta_L$:
\be\label{RLNL}
R(k,\eta)\equiv \frac{\la\delta_{NL}\delta_L^*\ra}{\la\delta_L\delta_{L}^*\ra}\ .
\ee
Clearly, $R$ can be obtained directly from the cross-correlation coefficient, $\rho_{NL,L}$ which is plotted in Figure~\ref{fig:rho}:
\be
R^2(k,\eta)=\frac{P_{NL}(k,\eta)}{P_L(k,\eta)}\rho_{NL,L}^2(k,\eta) \ .
\ee

%\begin{figure}[h!]
%\centering
%\includegraphics[width=5in]{rhos-eps-converted-to.pdf}
%\caption{\small We show the cross-correlation coefficient between the non-linear overdensity and the overdensity in linear theory and the ZA as a function of scale at $z=0$. The cross-correlation coefficient, $\rho_{NL,L}$ is very close to the non-linear propagator in RPT, despite their different definitions. Note that the Zel'dovich density field is well-correlated with the non-linear density field well into the mildly non-linear regime.} \label{fig:rho}
%\end{figure}

To understand the above split better, let us write it for the density field in the Zel'dovich approximation (ZA) as we did in Section~\ref{intro}. The Zel'dovich power spectrum, $P_Z$ can be written as:
\be
P_Z=R_Z^2P_L+P^Z_{MC}\ ,
\ee
with $R_Z$ defined in analogy with $R$ in eq.~(\ref{RLNL}) but with $\delta_{NL}$ replaced by the Zel'dovich density field, $\delta_Z$. The exact expression for the matter power spectrum, $P_Z(k)$, in the ZA is:
\be
\la \delta_Z(\bm{k})\delta_Z(\bm{k}')\ra&\equiv&\d(\bm{k}+\bm{k}')P_Z(k)\label{PZ}\\
\nonumber
&=&\d(\bm{k}+\bm{k}')\int \frac{d^3q}{(2\pi)^3} e^{-i\bm{k}\cdot \bm{q}}\exp\left[-\, k_i k_j \left(\delta_{ij}\sigma_v^2-\Psi_{ij}(\bm{q})\right)\right]\ ,
\ee
where 
\be
\Psi_{ij}(\bm{q})\equiv\int d^3\xi \frac{\xi_i\xi_j}{\xi^4} P_L(\xi)e^{i\bm{q}\cdot\bm{\xi}}
\ee
 is proportional to the linear velocity correlation function. The velocity dispersion is proportional to $\sigma_v$, which is defined as $\sigma_v^2\equiv\Psi_{ii}(0)/3$.
 
At first order in $P_L$, we have $P_Z\approx P_L$ as expected. At second order, one can check explicitly that the large-scale modes affect $P_Z$ only through the density power spectrum, and not through the velocity power spectrum. Thus, the ZA is not affected by bulk flows at second order. This result can be extended to all orders in the overdensity \cite{1996ApJS..105...37S}.

If we follow \cite{matsubara} to recover the MLTP result, we should expand $P_Z$ in $\Psi_{ij}$, while keeping $\sigma_v^2k^2$ in the exponent. The first order result in this expansion becomes $\exp (-k^2\sigma_v^2)P_L$ which we can identify with $R_Z^2 P_L$ \cite{Valageas:2007ge}. The rest of the terms must be collected in $P_{MC}$. 

Therefore, the split above distributes the canceling contributions from the bulk flows (terms containing $\sigma_v(<k)$) between $R^2_ZP_L$ and $P_{MC}^Z$ (see also Section~\ref{intro}). The expansion in $\Psi$ above is in the parameter $\sim k^2\sigma_v^2$ which is $\sim 1$ in the mildly non-linear regime. Thus, we can expect that $P^Z_{MC}$ is comparable to if not larger than $R_Z^2P_L$, which is indeed the case (see Figure~\ref{fig:Power}, keeping in mind that $R_Z$ and $R$ are almost equal).

%\begin{figure}[t!]
%\centering
%\includegraphics[width=5in]{FSpectra8Arrows2-eps-converted-to.pdf}
%\caption{\small Shown schematically are various matter power spectra at $z=0$ for $\Lambda$CDM. The power spectra are divided by a smooth BBKS \cite{BBKS} power spectrum with shape parameter $\Gamma=0.15$ in order to highlight the wiggles due to the BAO. The non-linear power spectrum (i.e. the ``exact'' power spectrum obtained from N-body simulations) is given by $P_{NL}$; the linear power spectrum by $P_L$; the part due to the ``memory of the initial conditions'' by $R^2P_L$; the power spectrum calculated in the Zel'dovich approximation by $P_Z$; the power due to the projection of the non-linear density field on the Zel'dovich density field by $\tilde R^2P_Z$. The green (long dashed) arrow  represents the power generated from mode-coupling due to both free-streaming and structure formation; while the blue (dot-dashed) arrow represents the mode-coupling power generated by structure formation alone. } \label{fig:Power}
%\end{figure}

One can apply the analysis above to the non-linear density field as well, noting that $R_Z$ is a very close approximation to $R$  \cite{Crocce:2005xz}. Thus, we can conclude that the split, eq.~(\ref{split}), is not optimal for analyzing $P_{NL}$. Therefore, for example, by choosing an initial power spectrum with spectral index $-3<n\leq -1$, one can arrange for $\sigma_v^2(<k)k^2$ to diverge, while the overdensity remains finite. In that limit, RPT and MLPT break down completely; while the split proposed below avoids these divergences, as equal-time statistics cannot depend on the large-scale bulk flows.

% A better approximation to $P_Z$ would be to expand in the quantity $\left(\delta_{ij}\sigma_v^2-\Psi_{ij}(\bm{q})\right)$ in eq.~(\ref{PZ}) which yields back $P_L$ at first order. Indeed $P_L$ is a much better approximation to $P_Z$ than $R^2P_L$. However, we would like to go beyond linear theory, and
So, instead of writing the above split around $P_L$, we perturb around the Zel'dovich approximation, which captures the effects of the bulk flows (see Section~\ref{intro}; and \cite{1996ApJS..105...37S} as well). Thus, we will keep $P_Z$ exact, and expand $P_{NL}$ around it:
\be\label{splitAroundZ}
P_{NL}=\tilde R^2(k,\eta)P_Z(k,\eta)+\tilde P_{MC}(k,\eta)\ ,
\ee
with
\be\label{RLNLZ}
\tilde R(k,\eta)\equiv \frac{\la\delta_{NL}\delta_Z^*\ra}{\la\delta_Z\delta_{Z}^*\ra}\ .
\ee
In \cite{HHpaper} it was shown that the splitting above emerges naturally when considering the statistical properties of CDM in phase-space. 

\section{Correcting for sample variance in N-body simulations}\label{sec:correcting}

The Zel'dovich density field and the non-linear density field are very well cross-correlated across the mildly-nonlinear regime (Figure~\ref{fig:rho}). Thus, one can treat them as simple rescalings of each other, the rescaling given by $\tilde R$.

As we discussed in Section~\ref{intro}, we expect $P_Z$ to capture most of the effects of the large-scale modes, which are the ones most affected by sample variance. Thus, let us assume for now that  $\tilde R$ is not affected by sample variance, and therefore does not depend on the particular realization of the random density field (see Section~\ref{whyWork}). 
%Then for scales where $\tilde P_{MC} $ is negligible, implying cross-correlation close to one, we can write\footnote{In principle $\tilde R$ in equations (\ref{splitAroundZ}) and (\ref{deltaRtilde}) must be different -- the former equation resulting after ensemble averaging the square of the latter. However, as we will argue $\tilde R^2-1$ is mainly due to large-scale modes. In the limit when those modes are much larger than the modes of interest, one can write an effective $\delta_{NL,\rm eff}\equiv\langle \delta_{NL}\rangle_{\Lambda}$ which gives the short-scale overdensity in a particular realization after ensemble averaging only over the large-scale modes (with $k$ less than a cutoff $\Lambda$). Thus, as written, eq.~(\ref{deltaRtilde}) should really be treated as an equation for this $\delta_{NL,\rm eff}$. However, $\tilde R^2-1$ is small compared to one, and therefore, for the qualitative arguments presented in this section we treat $\delta_{NL,\rm eff}$ and $\delta_{NL}$ as identical.}
We can then write:
\be\label{deltaRtilde}
\delta_{NL}(\bm{k},\eta)\approx \tilde R(k,\eta)\delta_Z(\bm{k},\eta)+\delta_{MC}(\bm{k},\eta)\ ,
%=\sqrt{\frac{P_{NL}(k,\eta)}{P_Z(k,\eta)}}\delta_Z(\bm{k},\eta)\ ,
\ee
where  $\delta_{NL}$ and $\delta_Z$ are calculated in the same realization, and $\delta_{MC}$ is defined such that 
$$
\la \delta_{MC}\delta_Z^*\ra =0\ .
$$
 Thus, for each simulation box of $\delta_{NL}(\bm{k},\eta)$, one has to run a corresponding Zel'dovich ``simulation'' with the same initial conditions to get the corresponding $\delta_Z(\bm{k},\eta)$.

Note that eq.~(\ref{deltaRtilde}) can be treated as a linear model for $\delta_{NL}(\delta_Z)$, with a non-random coefficient $\tilde R(k,\eta)$ and a random residual given by $\delta_{MC}$. The least-squares fit gives the following estimators (denoted with hats) for $\tilde R$ and $P_{MC}$:
\be
\hat R & = & \frac{\la \delta_{NL} \delta_Z^* \ra}{\la \delta_Z \delta_Z^* \ra}\ ,\\
\hat P_{MC} & = & \la \delta_{NL}\delta_{NL}^*\ra-\hat R^2 \la \delta_{Z}\delta_{Z}^*\ra\ .
\ee
Thus, we propose the following estimator for $P_{NL}$:
\be\label{restore}
\hat P_{NL}(k,\eta) = \hat R^2 P_Z + \hat P_{MC} = \left[\frac{\la \delta_{NL} \delta_Z^* \ra}{\la \delta_Z \delta_Z^* \ra}\right]^2 \bigg(P_Z-\la \delta_Z \delta_Z^* \ra \bigg)+\la \delta_{NL}\delta_{NL}^*\ra \ ,
\ee
where $P_Z$ is defined below (eq.~(\ref{P_Z_fixed})). The averages above denoted by the angular brackets are averages both over  the modes in each $k$-shell in each realization, \textit{and} over all simulation boxes. Therefore, the estimators above are manifestly unbiased in the limit of an infinite number of (finite in size) simulation boxes.

The Zel'dovich power spectrum $P_Z$ on the right-hand side of eq.~(\ref{restore}) is the exact theoretical power spectrum, $P_Z^T$ (we put a subscript $T$ to distinguish the power spectrum given in eq.~(\ref{PZ}) from the Nyquist-corrected $P_Z$ below), corrected for the truncation of $P_L$ at the Nyquist wave-vector, $k_N$. A good estimate for $P_Z$ is given by:
\be\label{P_Z_fixed}
P_Z(k,\eta)\approx P_Z^T(k,\eta) \exp\left(\frac{4\pi}{3}k^2\int\limits^\infty_{k_N}P_L(w,\eta)dw\right)\ ,
\ee
which corresponds to (\ref{PZ}) with a velocity dispersion, $\sigma_v^2$, evaluated up to $k_N$. Alternatively, one can compute $P_Z$ from an ensemble of initial conditions, since each Zel'dovich ``simulation'' is cheap to compute. The latter method is slower, but captures the effects of the cubical cells for initial conditions on a grid, or the effects from glass-like initial conditions. 

By applying eq.~(\ref{restore}) we will show that we can extract $P_{NL}(k,\eta)$ with a great accuracy even from a single N-body simulation, as long as we calculate the Zel'dovich density field using the same initial conditions used for the fully non-linear N-body simulation. 
As we will show below, one can even do better by applying eq.~(\ref{restore}) using the overdensity field obtained at second order in LPT (2LPT), instead of the Zel'dovich overdensity:
\be\label{restore2}
\hat P_{NL}(k,\eta)
= \left[\frac{\la \delta_{NL} \delta_{2LPT}^* \ra}{\la \delta_{2LPT} \delta_{2LPT}^* \ra}\right]^2 \bigg(P_{2LPT}-\la \delta_{2LPT} \delta_{2LPT}^* \ra \bigg)+\la \delta_{NL}\delta_{NL}^*\ra \ .
%\approx P_{2LPT}(k,\eta)\frac{\la\delta_{NL}(\bm{k},\eta)\delta^*_{NL}(\bm{k},\eta)\ra}{\la\delta_{2LPT}(\bm{k},\eta)\delta^*_{2LPT}(\bm{k},\eta)\ra} \ .
\ee
The exact $P_{2LPT}$ above can be computed from a large ensemble of 2LPT ``simulations'', which are cheap to compute.

In order to test the estimators for $P_{NL}$ we ran 10 simulations using GADGET-2 \cite{gadget} with $256^3$ particles each, in a box with side $L_{\mathrm{box}}=500\,h/$Mpc. The cosmological parameters  we used are (in the standard notation) as follows
\be
(\Omega_b,\Omega_{\mathrm{matter}},\Omega_\Lambda,h,n_s,\sigma_8)=(0.046,0.28,0.72,0.70,0.96,0.82)\ .
\ee
The initial conditions were set at a redshift of $49$, using the 2LPT code provided by \cite{2006MNRAS.373..369C}. The density field was calculated on a $512^3$ grid using Cloud-in-Cell (CIC) assignments, and the final power spectra were CIC corrected. 

\begin{figure}[h!]
\centering
\includegraphics[width=15cm]{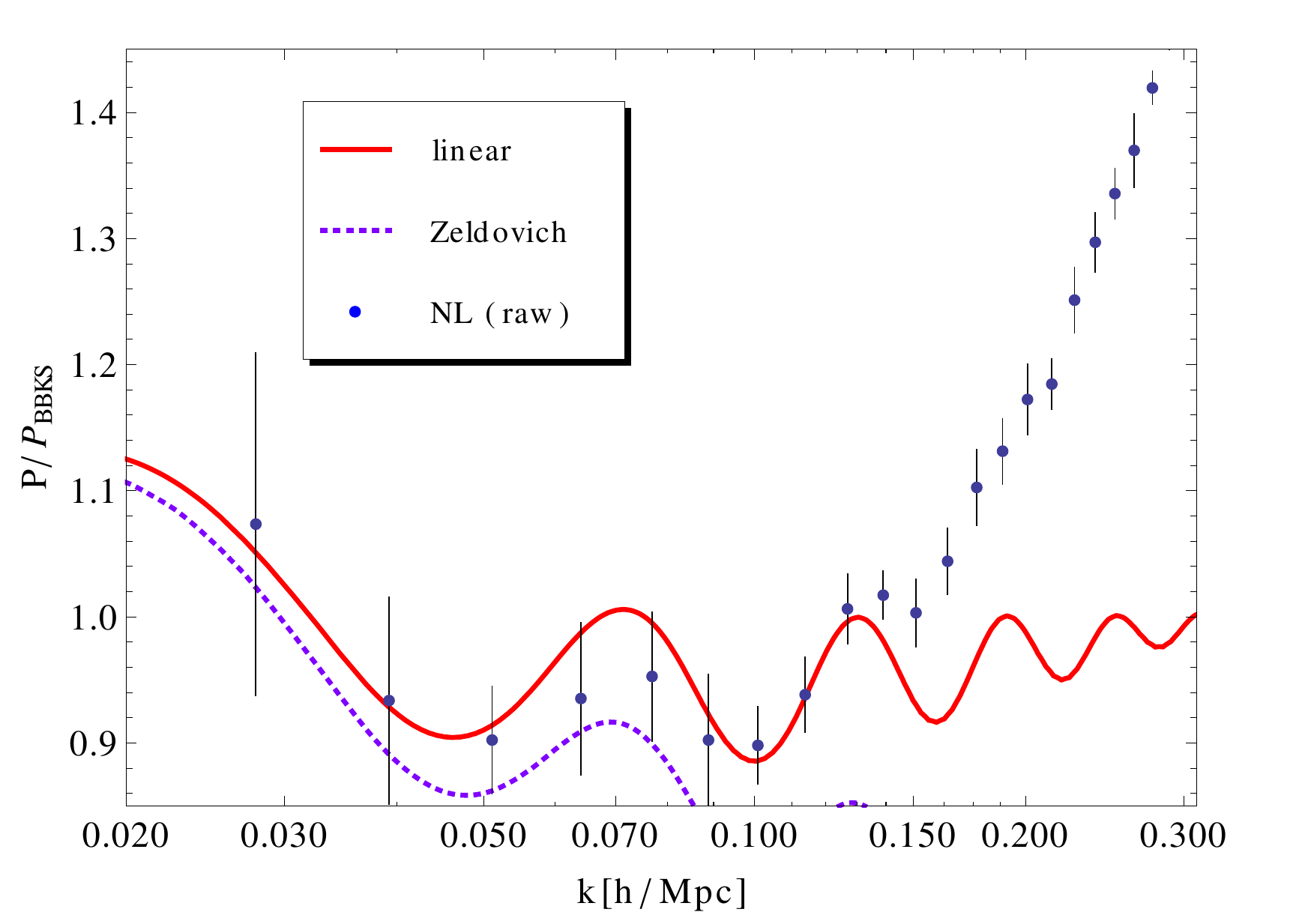}
\caption{\small We show the raw non-linear matter power spectrum at $z=0$ (with 2-sigma errorbars for the average $P_{NL}$) obtained from an ensemble of 10 N-body simulations. We also show the linear and Zel'dovich power spectra for comparison. All power spectra are divided by a smooth BBKS power spectrum to highlight the BAO wiggles. } \label{fig:RAW}
\end{figure}

\begin{figure}[h!]
\centering
\includegraphics[width=15cm]{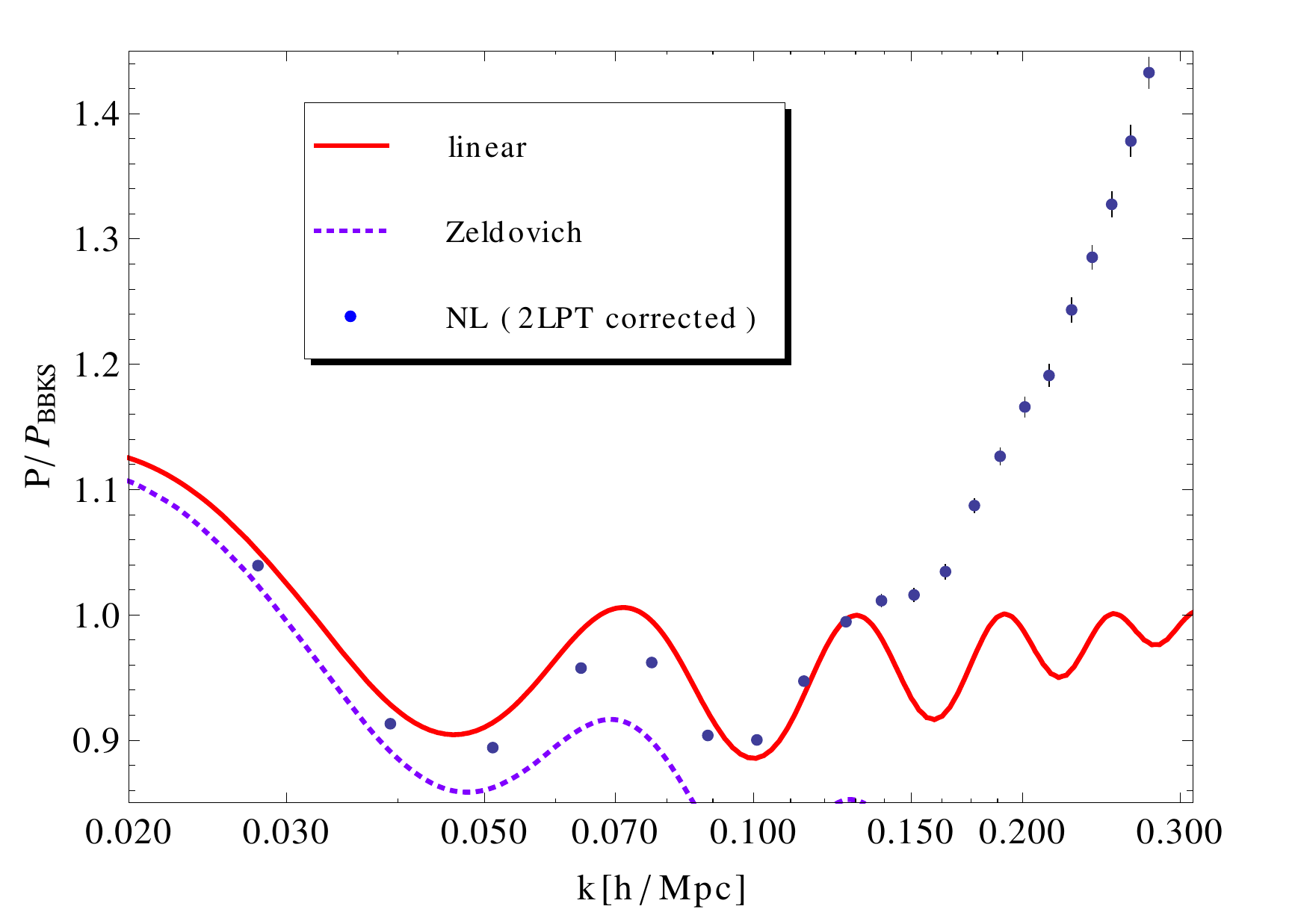}
\caption{\small The same as in Figure~\ref{fig:RAW} with the non-linear power spectrum calculated from the same 10 simulations but using the 2LPT estimator, eq.~(\ref{restore2}).} \label{fig:LPT}
\end{figure}

In Figure~\ref{fig:RAW} we show the non-linear power spectrum calculated directly  from the simulations.  Note the large errorbars on the average of $P_{NL}$ which are a direct consequence of sample variance. In Figure~\ref{fig:LPT} we plot the same quantities as in Figure~\ref{fig:RAW} obtained from the same 10 simulations but corrected using the 2LPT scheme, eq.~(\ref{restore2}). To do that, we used the following method: 
For each of the 10 N-body simulations, we ran a 2LPT ``simulation'' with the same initial conditions. To obtain $P_{NL}$ we used the 2LPT estimator, eq.~(\ref{restore2}), with the averages running over both all $k$-modes and all simulation boxes. The true $P_{2LPT}$ required for the 2LPT estimator was obtained from an average of 400 2LPT simulations. Then to calculate the errorbars, we calculated $P_{NL}$ using eq.~(\ref{restore2}) from each of the 10 realizations; took the rms of the resulting ten $P_{NL}$ estimates, and divided by $\sqrt{10}$ to obtain the 1-sigma error in the mean $P_{NL}$. 
The results are shown in Figure~\ref{fig:LPT}.
One can clearly see that the sample variance has been practically eliminated in the linear regime; and is significantly reduced in the mildly non-linear regime.

\begin{figure}[h!]
\centering
\includegraphics[width=15cm]{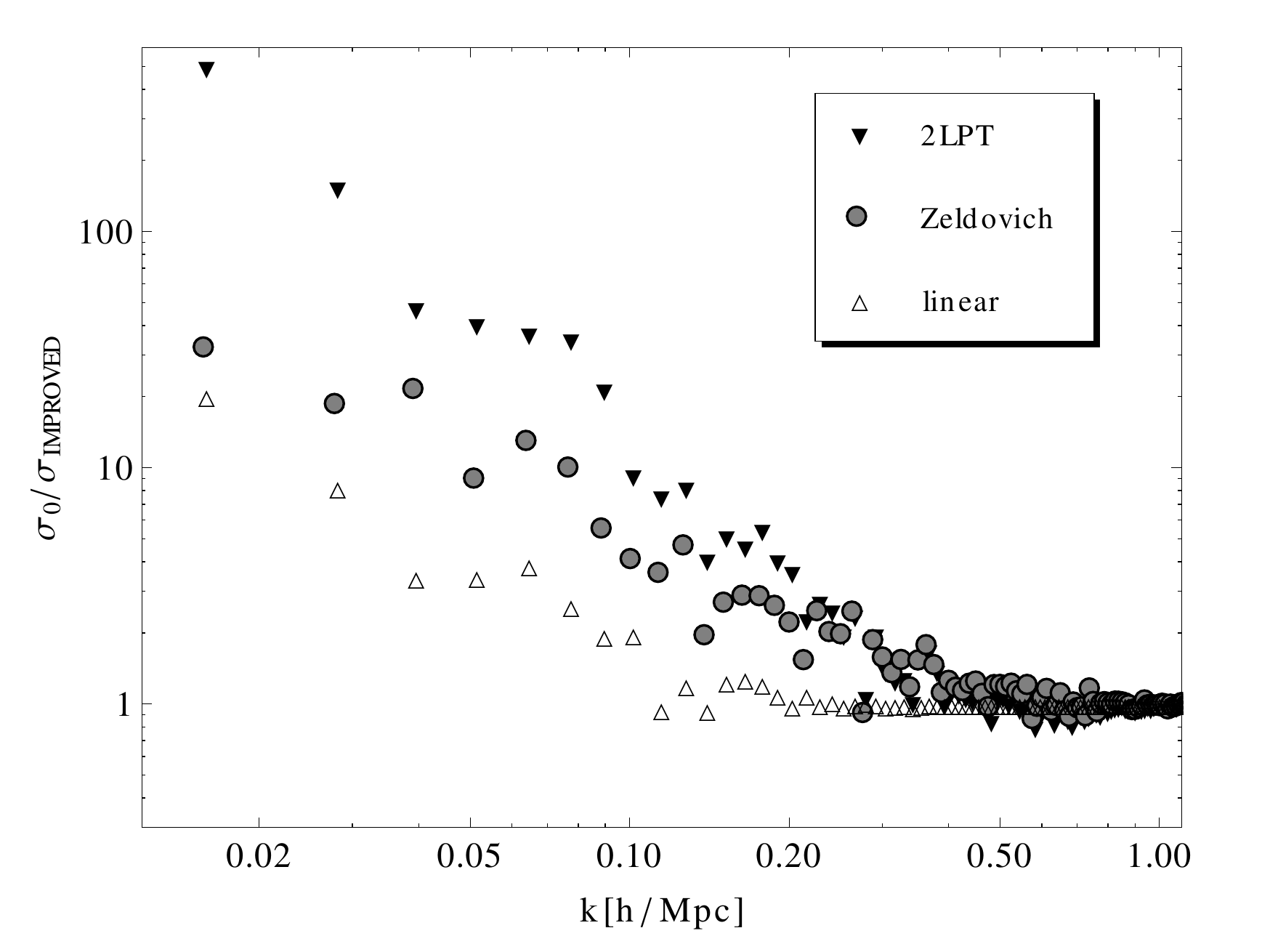}
\caption{\small We show the improvement in the error, $\sigma$, in determining $P_{NL}$ at $z=0$ using the restoration schemes discussed in the text: using eq.~(\ref{restore}) applied with 2LPT, the Zel'dovich approximation, and linear theory. We denote the error in $P_{NL}$ as obtained directly from the N-body simulations as $\sigma_0$. } \label{fig:Sigma}
\end{figure}

In Figure~\ref{fig:Sigma} we show the amount of improvement in the errors in determining $P_{NL}$ using eq.~(\ref{restore}) applied with 2LPT (i.e. eq.~(\ref{restore2})), the Zel'dovich approximation, and linear theory (by replacing the quantities evaluated in the ZA in eq.~(\ref{restore}) by their counterparts in 2LPT and linear theory). In the BAO region ($0.05h/$Mpc$<k<0.15h/$Mpc) the 2LPT estimator results in a reduction of the errorbars by about 5-50 times (depending on the scale) compared to the standard result. This implies that to achieve the same errors in the power spectrum in that region, the necessary computational time is reduced by a factor of $\sim30-2000$. 
Linear theory produces much worse improvement than LPT due to the fact that the cross-correlation $\rho_{NL,L}$ deviates from 1 at much larger scales (Figure~\ref{fig:rho}; see also Section~\ref{whyWork}). As discussed in  Section~\ref{intro}, this in turn can be understood by the fact that unlike LPT, linear theory does not properly account for the random bulk motions, which decorrelate the linear and non-linear density fields.

Note that the improvements obtained in both 2LPT and the ZA are no longer significant above $k\sim 0.5\,h$/Mpc. This is the scale where both cross-correlations $\rho_{NL,Z}$ and $\rho_{NL,2LPT}$ become $\sim 0.9$ (see Figure~\ref{fig:rho} for $\rho_{NL,Z}$). Neither the ZA nor 2LPT can capture the small-scale dynamics correctly. Thus, $\tilde P_{MC}$ grows large beyond  $k\sim 0.5\,h$/Mpc, and the $P_{NL}$ estimators we proposed no longer result in error improvements. However, those scales are usually well within the simulation volumes, and thus sample variance is usually less of a problem. Moreover, those scales are irrelevant for the acoustic peak from the BAO, since the oscillations in the power spectrum are completely washed away at these scales.

\section{Why does the method work?}\label{whyWork}

We need to understand why the procedure described in the previous section works so well in accounting for sample variance. 
For example, one may wonder why linear theory works so much worse than the ZA, when it models $P_{NL}$ even better than $P_Z$ (Figure~\ref{fig:Power}). Moreover, we have to try to understand why 2LPT works so much better than the ZA.

%The reason is that splitting $P_{NL}$ using linear theory as done in RPT and MLPT violates one important assumption in the analysis of the previous section -- the assumption that the mode-coupling power is negligible at the mildly non-linear scales (see also  Section~\ref{intro}).

As long as the cross-correlation coefficient $\rho_{PT,NL}$ is close to 1 (here $PT$ stands for 2LPT, Zel'dovich or linear theory), $\delta_{NL}$ and $\delta_{PT}$ are simple rescalings of each other. Therefore, we can write $\delta_{NL}\approx R \delta_{PT}$ (where for brevity we drop the tilde from $\tilde R$ and $\tilde P_{MC}$ whenever used in conjunction with the subscript $PT$, irrespective of whether $PT$ stands for LPT or not) with a non-random $R$. Our $P_{NL}$ estimator boils down to $\hat P_{NL}\approx R^2 P_{PT}$ with $R^2\approx \la |\delta_{NL}|^2\ra/\la |\delta_{PT}|^2\ra$. Thus, the $P_{NL} $ estimator has vanishing sample variance since for $\rho_{PT,NL}\approx 1$, $R$ must be non-random; and $P_{PT}$ is non-random by construction.

\begin{figure}
  \centering
  \includegraphics[width=15cm]{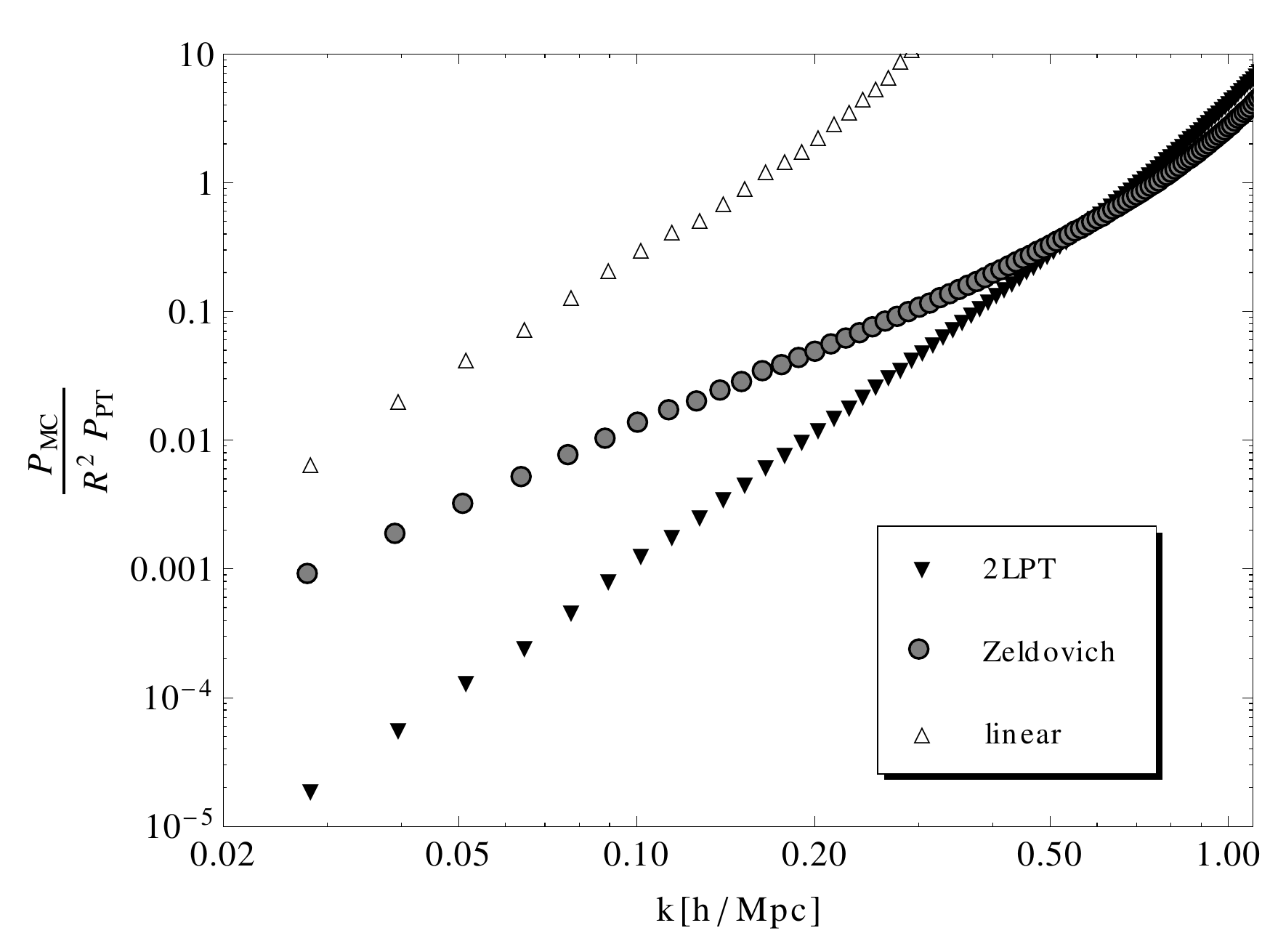}   \caption{\small We show the ratio of the two contributions to $P_{NL}$, eq.~(\ref{splitAroundZ}), depending on the choice of split (around the linear, Zel'dovich or 2LPT power spectrum) of $P_{NL}$. Note that the mode-coupling power is negligible at large scales, and becomes important only for scales for which $\rho_{PT,NL}$ deviates significantly from one.  }
  \label{fig:ratioAbs}
\end{figure}

\begin{figure}
  \centering
  \subfloat{\label{fig:gull}\includegraphics[width=0.495\textwidth]{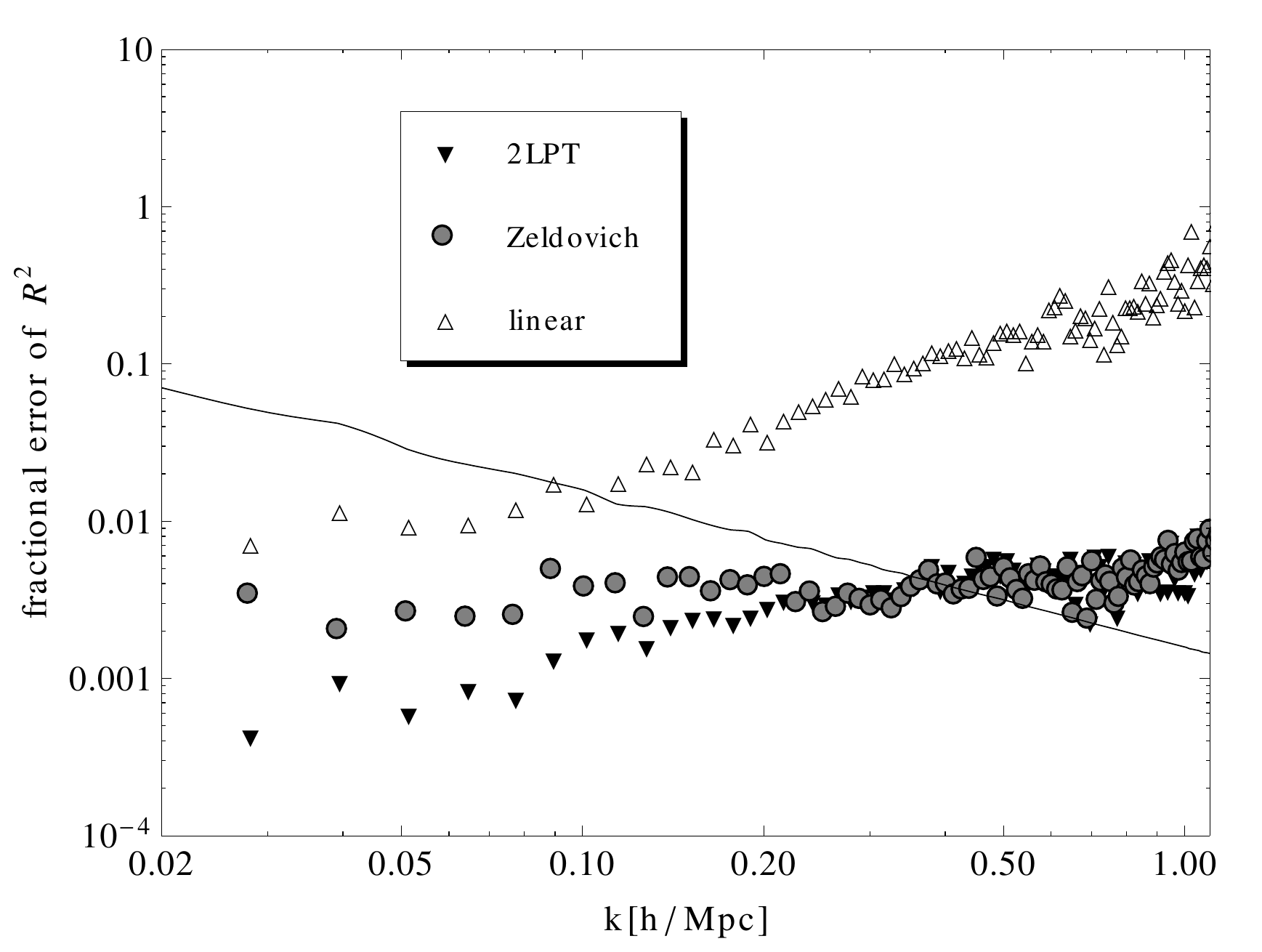}}     \hspace{0.002\textwidth}  \subfloat{\label{fig:tiger}\includegraphics[width=0.495\textwidth]{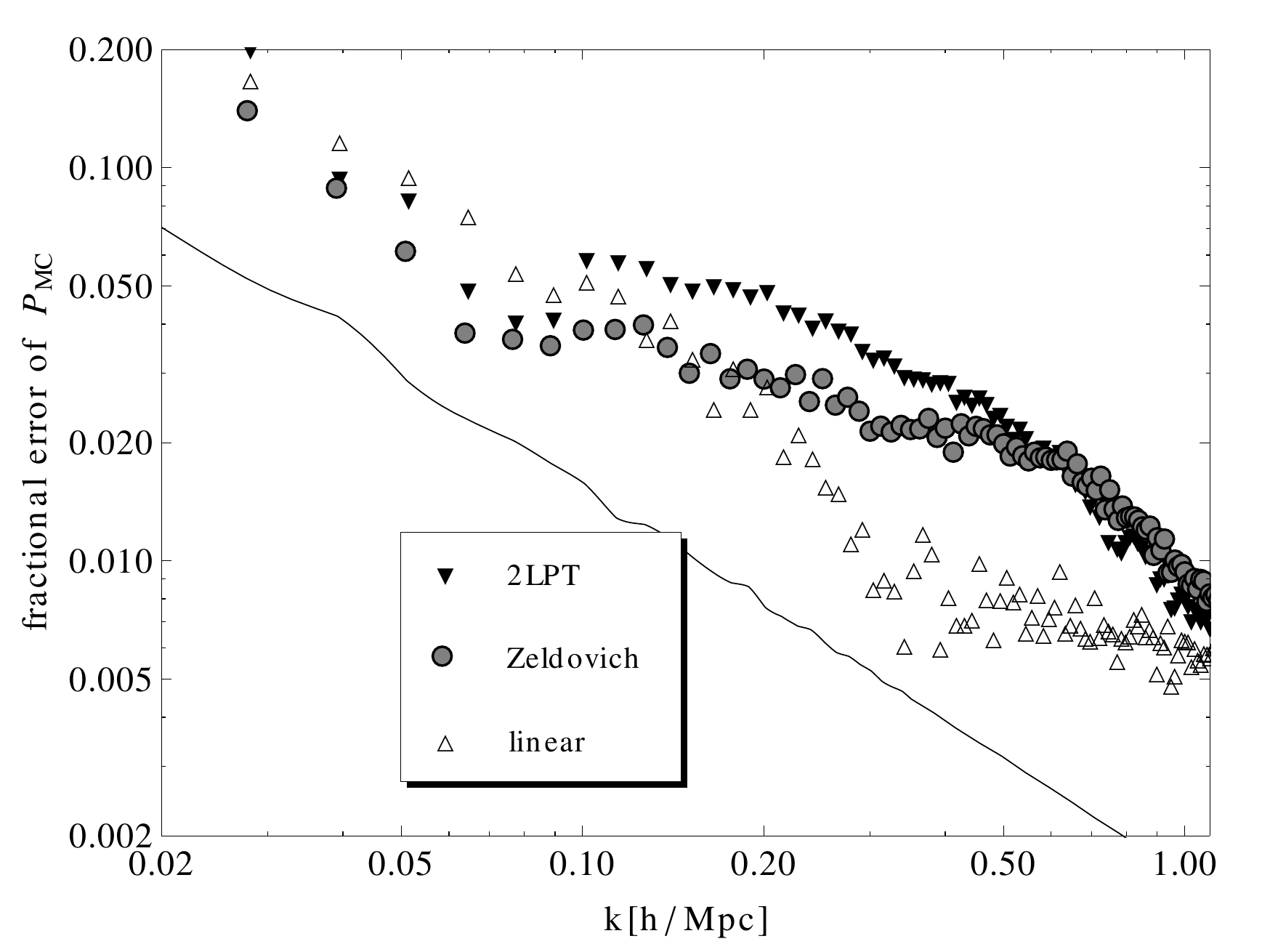}}
  \caption{\small We show the fractional error of $R^2$ and $P_{MC}$ (symbols), as well as the naive estimate of the error for Gaussian random variables, given by $\sqrt{2/N_k}$ (solid line).
  Note that the error of $R^2$ is greatly suppressed for $\rho_{PT,NL}\approx 1$.  }
  \label{fig:errorsMCR2}
\end{figure}

The above argument implies that when $\rho_{PT,NL} \approx 1$ we must have a negligible $P_{MC}$ \textit{and} the fractional error due to sample variance of $R^2 P_{PT}$ (which equals that of $R^2$, since $P_{PT}$ is the theoretical PT power spectrum) must be sub-Gaussian, i.e. less than $\sqrt{2/N_k}$, where $N_k$ is the number of modes in each $k$-bin from all simulations. To test the first prediction, in Figure~\ref{fig:ratioAbs} we show the ratio of the two terms entering in the split of $P_{NL}$, eq.~(\ref{splitAroundZ}), for the split applied around $P_L$, $P_Z$ and $P_{2LPT}$. One can see that at large scales, while $\rho_{PT,NL}\approx 1$, it is indeed $R^2P_{PT}$ that dominates over $P_{MC}$.  To test the second prediction, in Figure~\ref{fig:errorsMCR2} we plot the fractional errors of $R^2$ and $P_{MC}$. As expected, we see that  as long as $\rho_{PT,NL}\approx 1$ the fractional error of $R^2$ is much smaller than the variance for a Gaussian random variable, $\sqrt{2/N_k}$. This is not the case for $P_{MC}$ which has slightly super-Gaussian errors. However, $P_{MC}$ is negligible at large scales. 

We can conclude that for scales with $\rho_{PT,NL} \approx 1$ the overall error of $\hat P_{NL}$  must be greatly suppressed, which is indeed what we saw in the previous section. Moreover, since it is the large-scale modes which are most affected by sample variance, one can conclude that it is $R$ and $P_{PT}$, and not $P_{MC}$, that carry the information on the coupling to the large-scale modes. We will elaborate on this point further below.

As we saw, having $\rho_{PT,NL}\approx 1$ is crucial for the proposed split of $P_{NL}$, eq.~(\ref{splitAroundZ}), to work well. We can restate that condition as follows.
An adequate reconstruction method for $P_{NL}$ should rely on a split of $P_{NL}$ such that: 1) the mode-coupling power is small in the mildly non-linear regime, which implies that the approximate density field around which we expand (in the previous section that was the field in the ZA or 2LPT) is well correlated with the non-linear density field; 2) the power spectrum $P_{PT}$ of the approximate density field does not deviate much from $P_{NL}$ at the scales which have large sample variance. Both conditions above are satisfied in the ZA and 2LPT but not in linear theory. Linear theory has a large $P_{MC}$ in the mildly non-linear regime, which is a direct consequence of the $\rho_{L,NL}$ not being close to one. This explains why the split around $P_L$ is inferior to a split around the Zel'dovich or 2LPT power spectra.

Next, let us try to understand why 2LPT works so much better than the ZA. To do that we performed a series of numerical experiments. First we chose a scale $k_c=0.15\,h$/Mpc which is accessible using perturbation theory. We then ran a set of 10 simulations in which we cut off the high-k modes from the initial conditions (IC) by setting $P_{\mathrm{initial}}(k>k_c)=0$ -- we will refer to this set as the ``long'' set; and another set of 10 simulations in which we removed all large-scale modes from the IC: $P_{\mathrm{initial}}(k<k_c)=0$ -- we will refer to this set as the ``short'' set. We correct both sets using eq.~(\ref{restore2}) with the $\delta_{2LPT}$ calculated with the correspondingly band-passed initial conditions. 

\begin{figure}[t!]
\centering
\includegraphics[width=15cm]{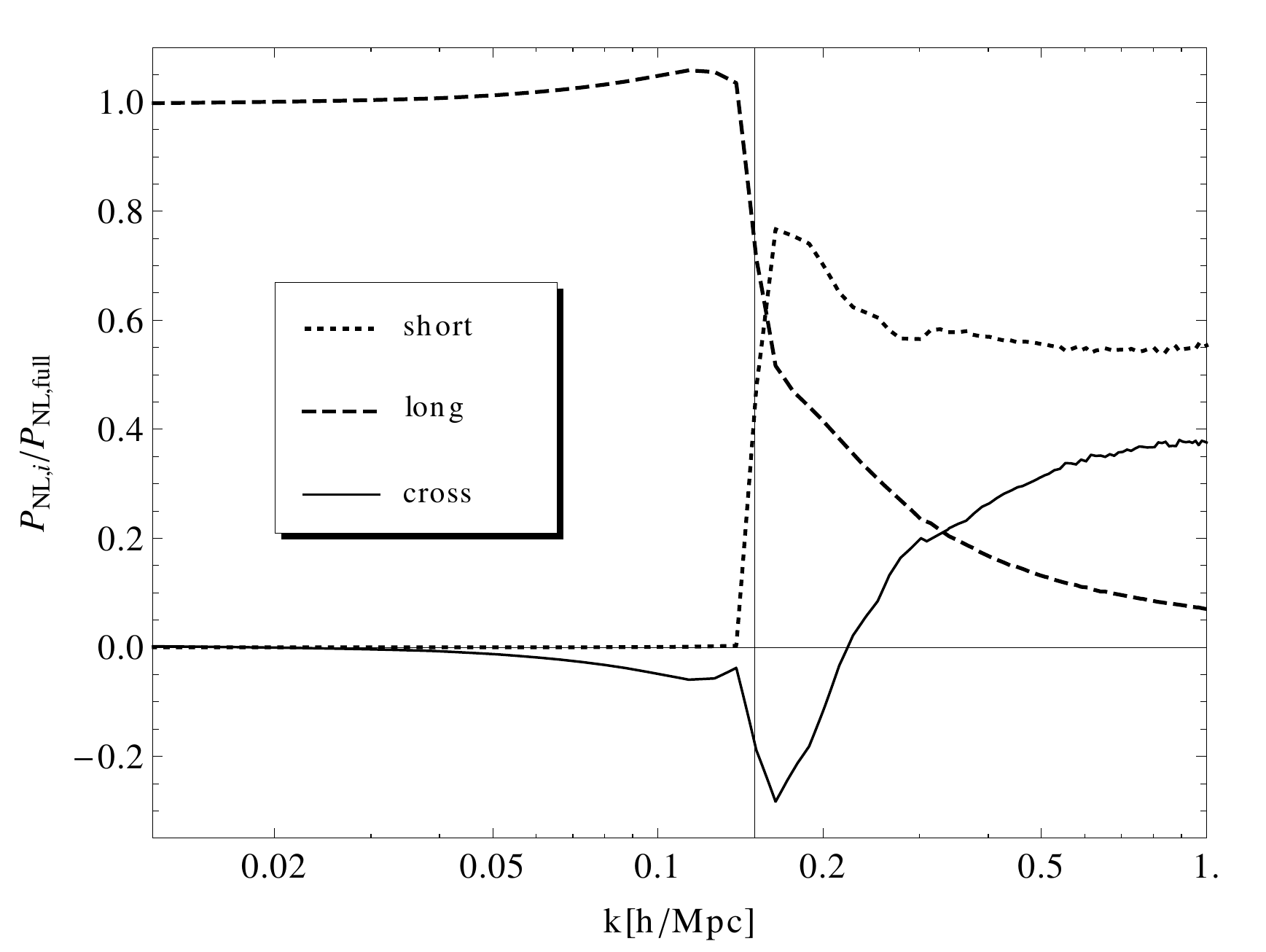}
\caption{\small We show the fraction of the full $P_{NL}$  at $z=0$ generated from the long and short IC, and the residual cross power spectrum. Errorbars are suppressed for clarity, but all curves are corrected for sample variance using eq.~(\ref{restore2}). The position of the cutoff $k_c$ is denoted by a vertical line.} \label{fig:contrNL}
\end{figure}

Mode coupling  generates power even at those scales, which had zero initial power. In Figure~\ref{fig:contrNL} we show the fraction of the full $P_{NL}$ which comes from the long and short IC, as well as the residual cross power, which arises from the coupling of the initially present short-scale and large-scale modes:
\be
P_{NL,\mathrm{full}}=P_{NL,\mathrm{long}}+P_{NL,\mathrm{short}}+P_{NL,\mathrm{cross}}\ ,
\ee
where a subscript ``full'' implies the quantity evaluated with the full (both large- and short-scale modes included) initial conditions, i.e. $P_{NL,\mathrm{full}}=P_{NL}$.

As one can see from Figure~\ref{fig:contrNL} practically no large-scale power is generated in the simulations with initially only short-scale modes. The reason is that mass and momentum conservation require that any power leaking from the small scales to the large scales should scale at least as $(k\sigma_v(>k_c))^4$ \cite{PeeblesBook}, where $\sigma_v(>k_c)\approx 2.8\,$Mpc$/h$ is the rms particle displacement due to the small scale ($k>k_c$) power. This scaling is modified to $\sim  (k/k_c)^2(k\sigma_v(>k_c))^4\approx (k/k_c)^6/32$ in Figure~\ref{fig:contrNL} since we have divided by $P_{NL}$ which scales as roughly $k^{-2}$ around $k_c$. This results in the sharp cutoff observed for $P_{NL,\rm{short}}$ for $k<k_c$. 

The cross power generated from the coupling of the initially long and short modes is again mostly localized at high $k>k_c$. This term tells us how the small-scales are influenced by the effects of the perturbed local $\Omega_{\mathrm{matter}}$ and by the tidal (and higher order)  interactions due to the presence of the large-scale modes.
%The cross power has an additional peak at around the cutoff scale, which tells us how neighboring (around $k_c$) modes are coupled.

The simulation with low-pass filtered IC  produces a fat tail at high $k>k_c$. This is a direct consequence of mode-coupling both due to structure formation and stream crossing.

For $k<k_c$ we can see that it is the low-pass filtered IC that dominate the true power spectrum, $P_{NL}$. Thus, we can conclude that the power in the mildly non-linear regime not captured by linear theory is due almost entirely to the coupling to the large-scale modes.

\begin{figure}[t!]
\centering
\includegraphics[width=15cm]{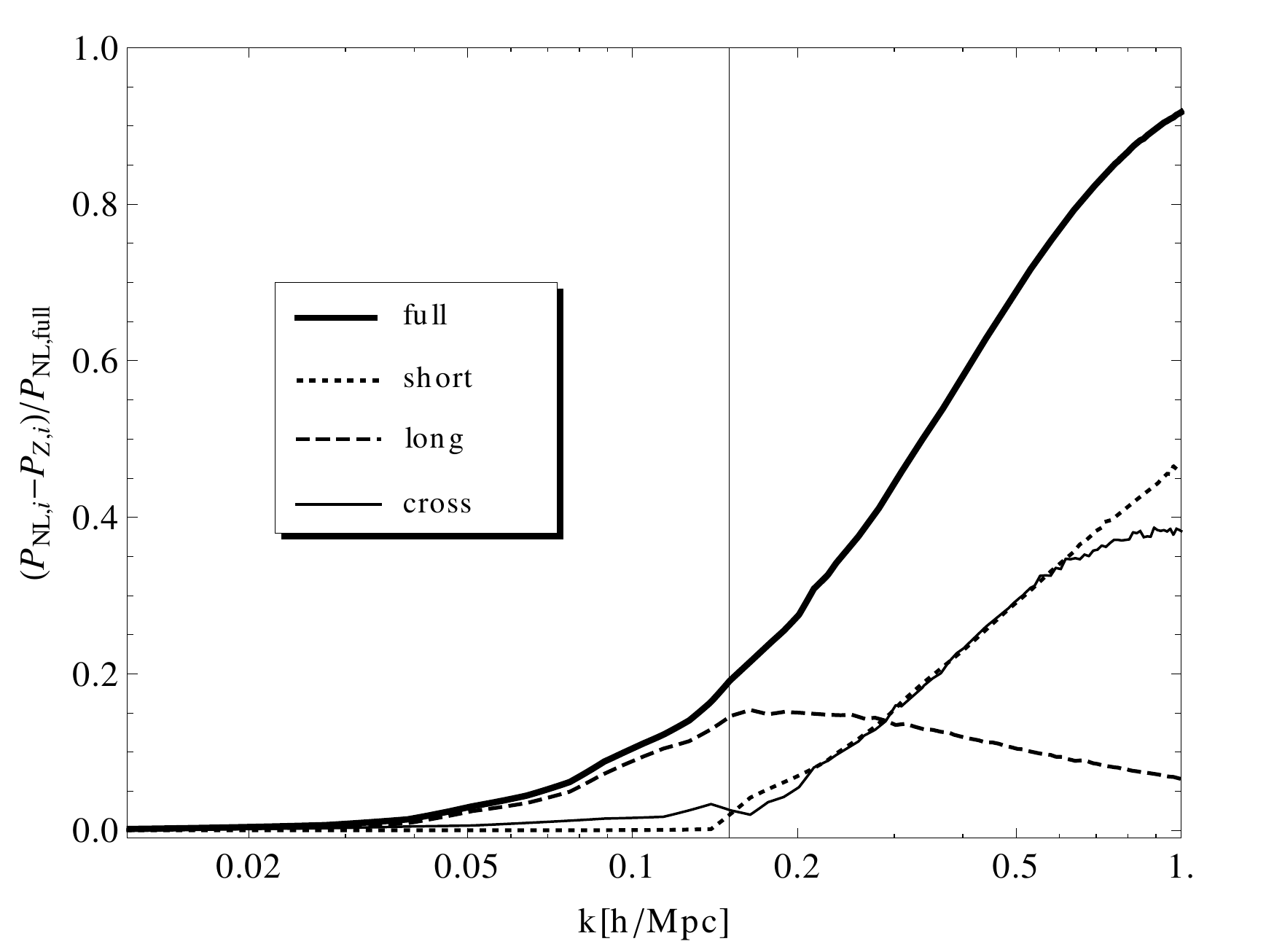}
\caption{\small We show the fractional contributions to the full $P_{NL}$ at $z=0$ which are not accounted for by the ZA: those contributions generated from the long and short IC, and the residual cross power spectrum. Errorbars are suppressed for clarity, but all curves are corrected for sample variance using eq.~(\ref{restore2}). The position of the cutoff $k_c$ is denoted by a vertical line.} \label{fig:contr}
\end{figure}

\begin{figure}[h!]
\centering
\includegraphics[width=15cm]{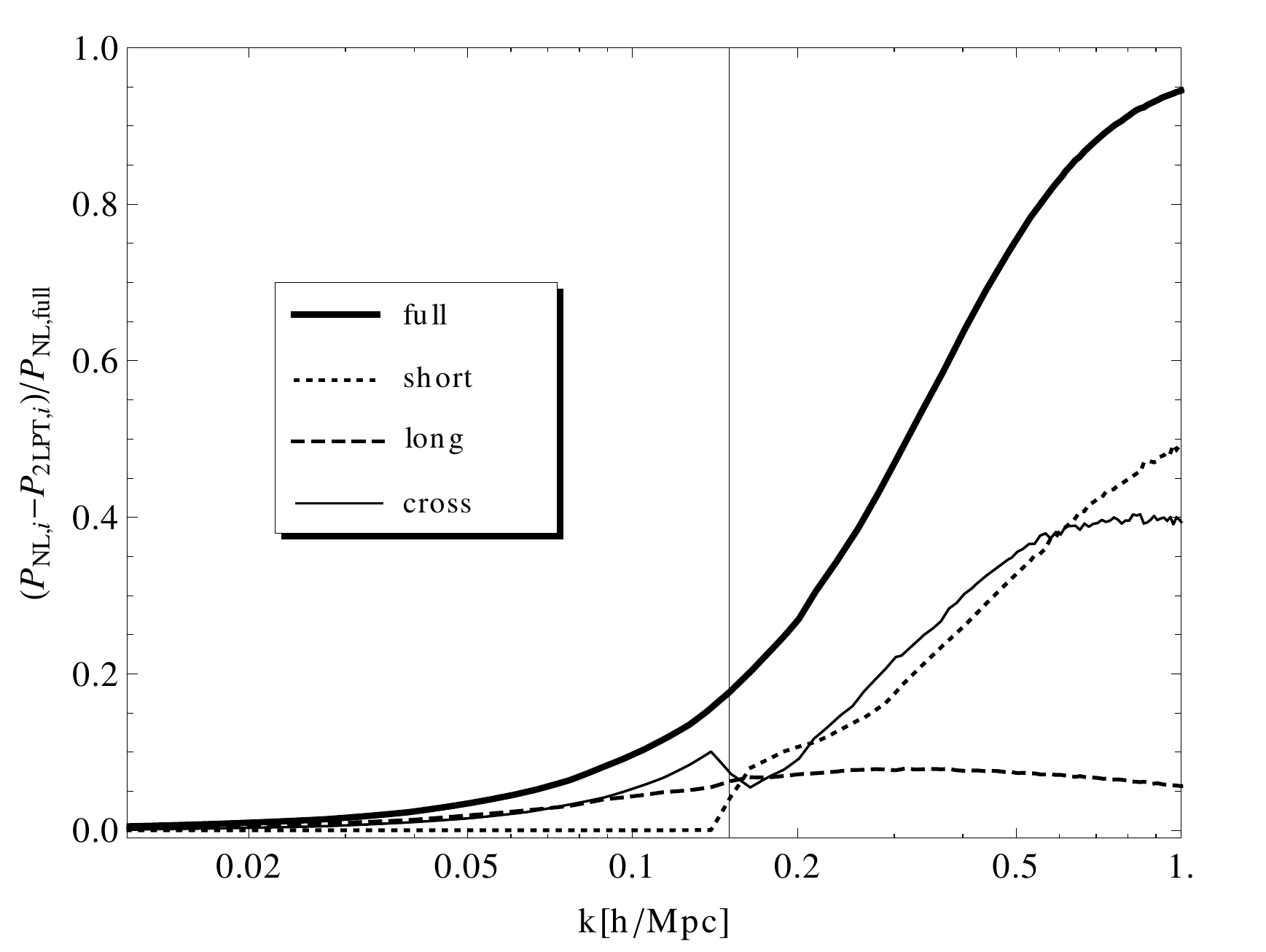}
\caption{\small As in Figure~\ref{fig:contr} but for the the fractional contributions to the full $P_{NL}$ which are not accounted for by 2LPT.} \label{fig:contr2LPT}
\end{figure}

Since the ZA and 2LPT satisfy both requirements above for an adequate reconstruction method, they should account for much of the large-scale power in $P_{NL}$, as well as the peak in the cross power at $k_c$, which at these scales should mostly be due to large-scale flows. 
To investigate this further, in Figures~\ref{fig:contr} and \ref{fig:contr2LPT}  we plot the different contributions, $i$, to the full non-linear power which are not accounted for by the Zel'dovich/2LPT approximation: $(P_{NL,i}-P_{Z/2LPT,i})/P_{NL,\mathrm{full}}$. For negligible $\tilde P_{MC}$ this quantity reduces to
\be\label{R2m1}
\frac{P_{NL,i}-P_{Z/2LPT,i}}{P_{NL,\mathrm{full}}}\approx \frac{\tilde R^2_i-1}{\tilde R^2_{\mathrm{full}}}\frac{P_{Z/2LPT,i}}{P_{Z/2LPT,\rm{full}}}\ ,
\ee
where a subscript ``full'' implies the quantity evaluated with the full initial conditions as before; and $\tilde R$ is defined either with respect to the ZA or 2LPT.

Comparing Figures~\ref{fig:contr} and \ref{fig:contr2LPT} to Figure~\ref{fig:contrNL} we can see that the peak of the cross power at $k\sim k_c$ is well-accounted for by LPT (both the ZA and 2LPT). So, indeed LPT captures quite well the cross-correlations between nearby large-scale modes.  The residual $\tilde R^2-1$ is due mostly to large-scale modes in the ZA (for $k\lesssim k_c$), while $P_{2LPT}$ captures better the coupling between the large-scale modes. Since it is precisely the large-scale modes which suffer most from sample variance, this explains why 2LPT is much better than the ZA in correcting for sample variance (see Figure~\ref{fig:Sigma}).

\section{Modeling the transfer function in the mildly non-linear regime}\label{sec:model}

In the previous sections we demonstrated the advantages of splitting the non-linear power spectrum as in eq.~(\ref{splitAroundZ}). Since the mode-coupling term, $\tilde P_{MC}$, is relatively small in the mildly non-linear regime, we can think of $\tilde R(k,\eta)$ as a transfer function transforming the power spectrum in the ZA to the true $P_{NL}$. 
%If a model for this transfer function is successfully constructed, then one will effectively solve the mildly non-linear regime. 
In this section we present a simple physically motivated model for the mildly non-linear regime from which we obtain a good approximation for $\tilde R$. In this section we will work only with $\tilde R$ as written in  eq.~(\ref{splitAroundZ}), i.e. with respect to the ZA, not 2LPT.

From Figure~\ref{fig:contr} and eq.~(\ref{R2m1}) one can see that $\tilde R^2-1$ is dominated by large-scale modes. Those affect the smaller scales by rescaling the local matter density (i.e. producing perturbations to the local $\Omega_{\rm matter}$); through tidal effects; and through higher order (in a multipole expansion) corrections. We will concentrate on the leading corrections, which correspond to those coming from the perturbed local $\Omega_{\rm matter}$. These perturbations lead to a perturbed local growth factor, whose effect is analyzed in \cite{baldauf}. These authors split the density field ($\delta=\delta_\Lambda+\delta_s$) into a large-scale ($\delta_\Lambda$) and a short-scale ($\delta_s$) component and find that the large-scale modes modify the short-scale modes as follows (in configuration space):
\be\label{deltaSNL}
\delta_{s,NL}\approx\delta_{s,L}\left(1+\frac{34}{21}\delta_{\Lambda,L}+\frac{341}{189}\delta_{\Lambda,L}^2\right)
\ee
plus corrections which are higher order in $\delta_{\Lambda,L}$ (for $k_{\rm short}\gg k_{\rm long}$, where the $k$'s correspond to $\delta_{s}$ and $\delta_{\Lambda}$, respectively). As usual above we used a subscript $L$ to denote quantities in linear theory, while $NL$ denotes the ``true'' result. The rescaling above due to the large-scale modes comes exactly from the corrections to the local growth factor from the slightly changed $\Omega_{\rm matter}$. Assuming a sharp $k$-cutoff for splitting $\delta$ into $\delta_s$ and $\delta_\Lambda$, the resulting short-scale power spectrum is 
\be\label{Pnlshort}
P_{NL}\approx P_{L}\left(1+\frac{8242}{1323}\,4 \pi\int\limits_0^\Lambda P_L(k)k^2 dk\right)=P_{L}\left(1+\frac{8242}{1323}\,\delta^2(<\Lambda)\right)\ ,
\ee
where we used that $k_{\rm short}\gg k_{\rm long}$; $\Lambda$ is a large-scale cutoff; and the linear short-scale and large-scale modes are independent: $\la\delta_{s,L}\delta_{\Lambda,L}\ra= 0$. As before $\delta^2(<\Lambda)$ is the density variance integrated up to $\Lambda$.

\begin{figure}[t!]
\centering
\includegraphics[width=5in]{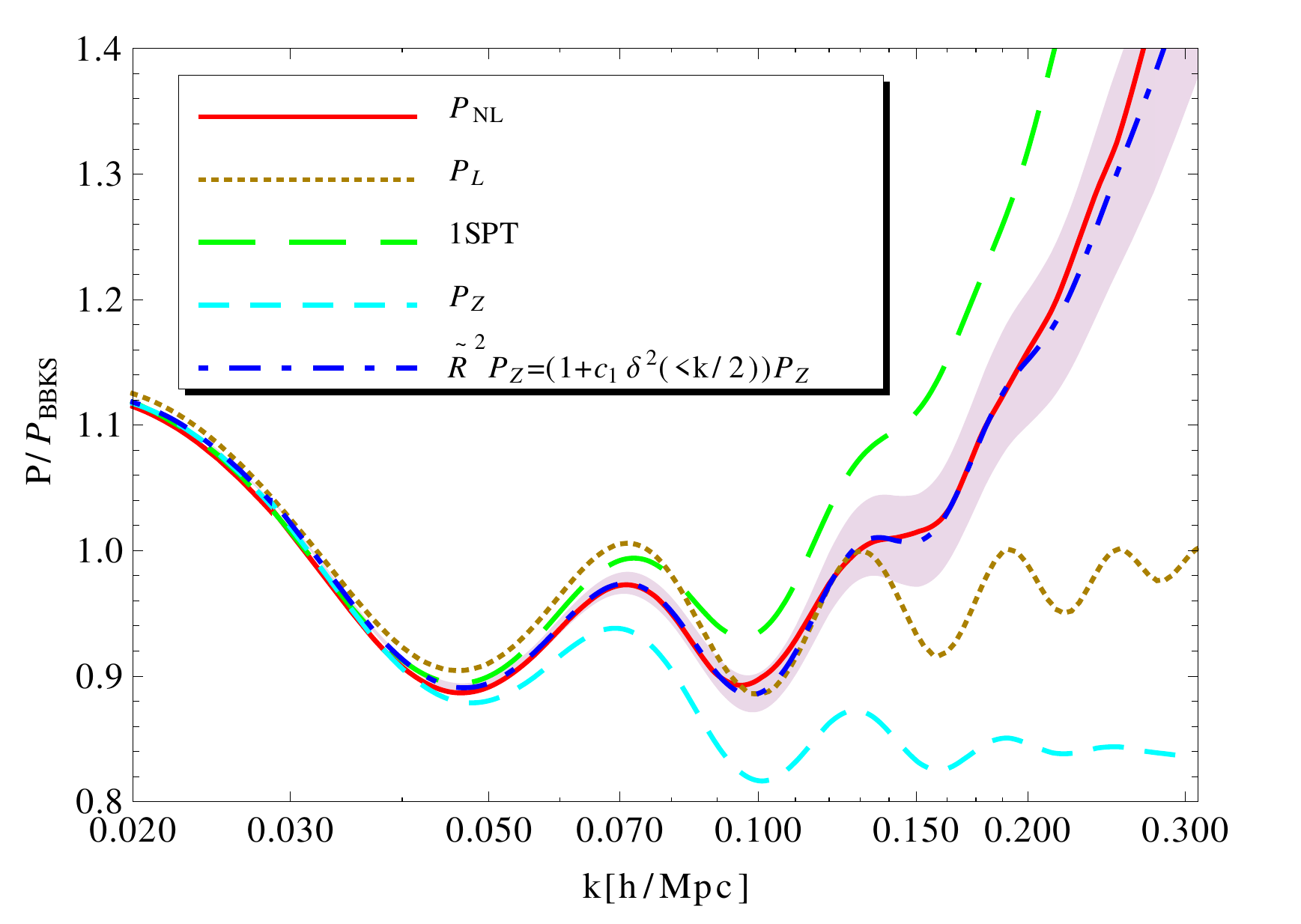}
\caption{\small Here we show various matter power spectra at $z=0$ for $\Lambda$CDM. As before, the power spectra are divided by a smooth BBKS power spectrum. The non-linear power spectrum is given by $P_{NL}$; the linear power spectrum by $P_L$; the 1-loop result in SPT is denoted by 1SPT; the power spectrum calculated in the Zel'dovich approximation is $P_Z$. Our formula for the power spectrum in the mildly non-linear regime results in a power spectrum given by the curve denoted with $\tilde R^2 P_Z$ (eq.~(\ref{PnlshortZ})), with $c_1=2950/1323\approx2.23$. The shaded region is bounded with the $P_{NL}$'s obtained from (\ref{PnlshortZold}) with $\Lambda=0.9 k/2$ (lower bound) and $\Lambda=1.1 k/2$ (upper bound).} \label{fig:MNLresult}
\end{figure}

In order to recover $\tilde R$, we should rewrite the above equation using $P_Z$, where for brevity we drop the subscript $T$ from $P_Z^T$. However, the ZA already captures some of the effects from $\delta_\Lambda$. By expanding $P_Z$ entering in $P_{NL}\approx\tilde R^2 P_Z$ in powers of $\delta^2(<\Lambda)$ and matching the resulting $P_{NL}$ to the short-scale power in eq.~(\ref{Pnlshort}) we obtain (see Appendix~\ref{appA})
\be\label{PnlshortZold}
P_{NL}(k,\eta)\approx P_{Z}(k,\eta)\left(1+\frac{2950}{1323}\,\delta^2(<\Lambda)\right)\ .
\ee

Note that the above equation is derived from the following expression relating the short-scale non-linear overdensity and the overdensity in the ZA valid for equal-time statistics (see Appendix~\ref{appA}):
\be\label{sNLZ}
\delta_{s,NL}=\delta_Z\left(1+\frac{2}{7}\delta_{\Lambda,L}+\frac{59}{189}\delta_{\Lambda,L}^2\right)\ .
\ee
This equation is our model for the overdensity in the mildly non-linear regime. It is correct to second order in $\delta_{\Lambda,L}$ (for $k_{\rm short}\gg k_{\rm long}$). However, when using the above expression one needs to keep in mind that $\la \delta_Z\delta_{\Lambda,L}\ra\neq 0$, unlike $\la\delta_{s,L}\delta_{\Lambda,L}\ra= 0$.

\begin{figure}[h!]
\centering
\includegraphics[width=5in]{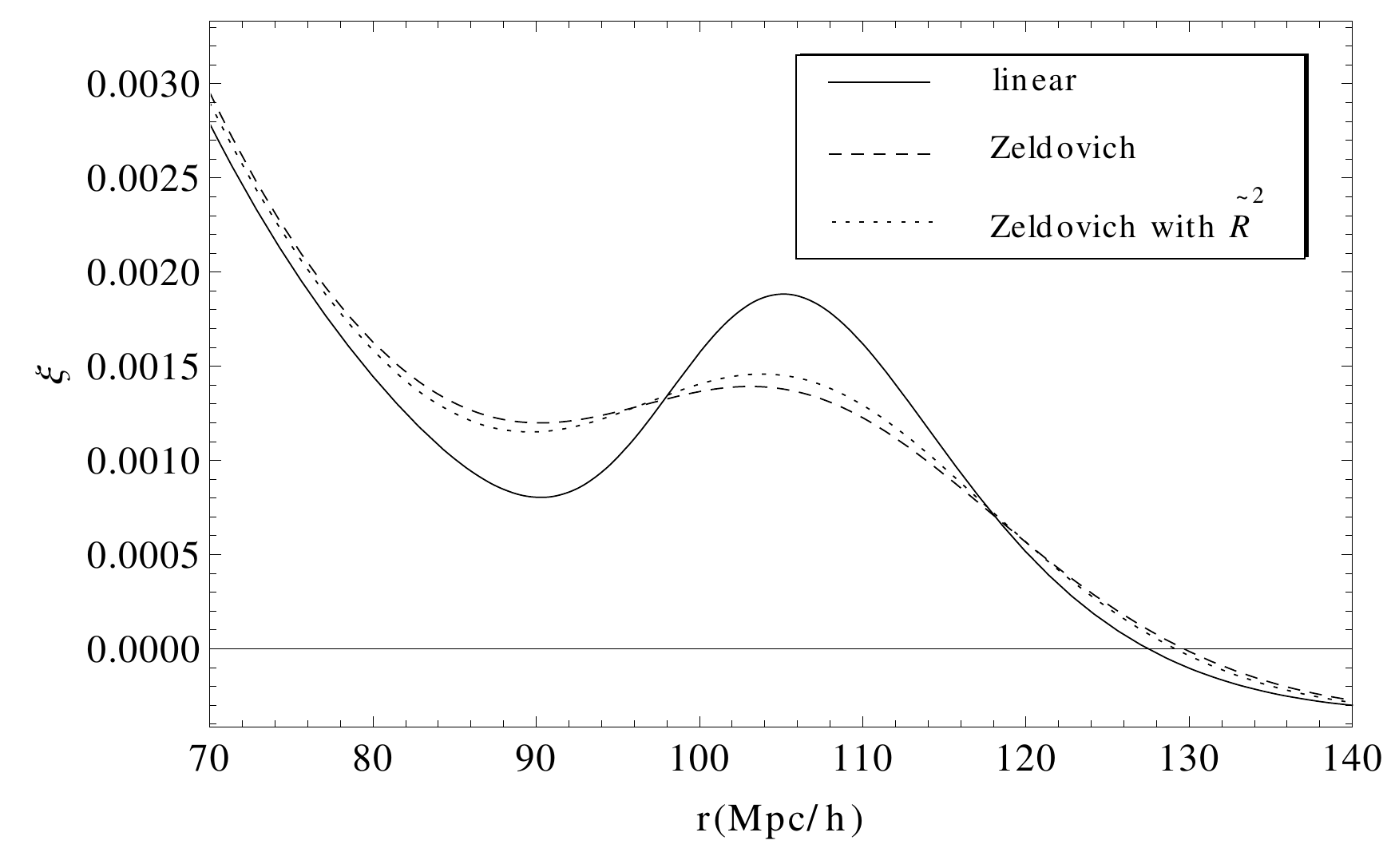}
\caption{\small Here we show the 2-pt function, $\xi$, at $z=0$ for $\Lambda$CDM. One can see the acoustic peak at $\approx 105\,$Mpc$/h$. The dotted curve is the $\xi$ corresponding to $\tilde R^2 P_Z$ with $\tilde R$ given by our model, eq.~(\ref{PnlshortZ}).} \label{fig:MNLresultXi}
\end{figure}

The above results surely break down for $\Lambda>k/2$, since then two large-scale modes can couple to directly produce a short-scale mode. Thus, we choose $\Lambda=k/2$:
\be\label{PnlshortZ}
P_{NL}(k,\eta)\approx P_{Z}(k,\eta)\left(1+\frac{2950}{1323}\,\delta^2(<k/2,\eta)\right)\ ,
\ee
from which one can read off $\tilde R^2$ as the term in the brackets.
 We plot this result in Figure~\ref{fig:MNLresult}. For reference we also give the corresponding 2-pt functions in real space in Figure~\ref{fig:MNLresultXi}.

From Figure~\ref{fig:MNLresult} one can see that the expression for the matter power spectrum, eq.~(\ref{PnlshortZ}), reproduces the true power spectrum to 1\% accuracy for $k\lesssim 0.2\,h/$Mpc. Thus our simple model is working much better than 1-loop SPT and has about the same level of accuracy as 2-loop RPT in the mildly non-linear regime \cite{Crocce:2007dt}. 

One may argue that the $(1/2)$ in $\delta^2(<k/2)$ in eq.~(\ref{PnlshortZ}) should really be treated as a fudge factor. Since in real space the acoustic peak is shifted by about 1\% from the result in linear theory, we varied the coefficient of (1/2) to produce a comparable shift. We found that deviations by as much as $50$\% are needed.\footnote{Varying the coefficient in front of $\delta^2(<k/2)$ by 50\% results in even smaller shifts in the acoustic scale.} These result in wildly different power spectra (see Fig.~\ref{fig:MNLresult} where with the shaded region we show the result from a 10\% change in the coefficient of (1/2)). Thus, barring any systematics, the model behind eq.~(\ref{PnlshortZ}) gives a robust estimate for the acoustic scale and the mildly non-linear power spectrum. Furthermore, one may be able to apply our model for the reconstruction of the BAO peak in an analogous way to \cite{2009PhRvD..79f3523P}.

Before we conclude, a word of caution is in order. In order to simplify the calculations in this paper, we assumed that the large-scale overdensity is spherically symmetric, eq.~(\ref{largeS}), and not a plane-wave as it should be. This assumption enters in the derivation of both $\delta_Z$, eq.~(\ref{deltaZT}), and $\delta_{s,NL}$, eq.~(\ref{deltaSNL}). In principle, this assumption can lead to problems. And indeed eq.~(\ref{PZPL}) tells us that $P_Z>P_L$, which is clearly not the case (e.g. Fig.~\ref{fig:MNLresultXi}). However, we use $P_L$ only for convenience in the intermediate results, hoping that any mistake we make, affects $P_Z$ and $P_{NL}$ equally (because of the strong correlation between $\delta_Z$ and $\delta_{NL}$), and therefore may cancel in the final result, (\ref{PnlshortZold},\ref{PnlshortZ}). Indeed our numerical results confirm this intuition. Clearly, if one wants to do better, one can repeat the calculation with a plane-wave large-scale mode. However, we leave that for future work, noting that our analytical formula (\ref{PnlshortZ}) gives rather robust results as it is.

Therefore, we can conclude that the CDM particle dynamics in the mildly non-linear regime is well approximated by the dynamics in the Zel'dovich approximation with perturbed growth factor. These perturbations to the local growth factor result from the perturbed local matter density caused by the presence of the large-scale modes.

\section{Summary}\label{sec:summary}

In this paper we used LPT to investigate the behavior of modes in the mildly non-linear regime. We find that their behavior is affected mainly by neighboring (in $k$) modes and by larger-scale modes. Much-smaller-scale modes are isolated and have almost negligible back-reaction. At the same time large-scale modes affect the mildly non-linear scales only by changing the local matter density or through subdominant tidal effects. 
%One should note that some of the contributions coming from the change in the local density are already captured by the ZA.
%At the same time larger-scale modes coherently move the mildly non-linear scales, which leads to no observable effects. 
%The overall density perturbation caused by larger-scale modes is captured by the ZA, since it simply changes the local growth factor, as it corresponds to a locally different $\Omega_{\rm matter}$. 

We find that the effects of the neighboring modes and large-scale modes are well modeled by LPT (but see below), which takes into account the effects of the bulk flows. Thus, unlike linear theory, the cross-correlation between the LPT overdensity and the non-linear overdensity is close to one in the mildly non-linear regime. We also find that the ratio (which approximately gives the square of the response (transfer) function, $\tilde R$) of the true power spectrum and the power spectrum in the ZA is nearly independent of realizations, which is due to the fact that the behavior of the large-scale modes, which are the ones with the large sample variance, is well-captured by LPT in the mildly non-linear regime.

Using the above results, we developed a straightforward method for correcting the results from N-body simulations for sample variance in the linear and mildly non-linear regimes. The method has no free parameters and introduces about an order of magnitude improvement in the errors in determining the true matter power spectrum.
The method can be used to speed up the scanning of the cosmological parameter space by an order of magnitude or more for observables in the (linear and) mildly non-linear regime. 

We then constructed a physically motivated model for the mildly non-linear regime, which treats the effects of the bulk flows correctly. The model is the same as the Zel'dovich approximation but with a perturbed local growth factor. Those perturbations are caused by the large-scale modes changing the local $\Omega_{\rm matter}$ -- an effect which (to first order in the large-scale power) is not completely captured by the ZA and 2LPT.

Using this model, we gave an approximate expression for the transfer function, $\tilde R$, which relates the power spectrum in the ZA, and the true power spectrum. The resulting power spectrum models the true power spectrum with an accuracy of $\lesssim1\%$ in the mildly non-linear regime, and gives robust estimates for the BAO scale.

\appendix
\section{Deriving equation (\ref{PnlshortZold})} \label{appA}

In this section we derive equation (\ref{PnlshortZold}). To do that let us start by writing down the expression for the Eulerian comoving coordinates $\bm{x}(\bm{q},\eta)$ of a particle in the ZA, which are a function of the particle's Lagrangian coordinates $\bm{q}$ and $\eta$:
\be
\bm{x}(\bm{q},\eta)=\bm{q}+D(\eta)\bm{s}(\bm{q})\ ,
\ee
where $D$ is the growth factor, and $\bm{s}$ is an irrotational displacement field given by
\be
-D(\eta) \nabla_{\bm{q}}\cdot \bm{s}(\bm{q})=\delta_L(\bm{q},\eta)\ .
\ee
The overdensity in the ZA, is then given by ($k\neq 0$):
\be
\delta_Z(\bm{k},\eta)=\int \frac{d^3x}{(2\pi)^3} e^{-i\bm{x}\cdot\bm{k}}  \int d^3 q\delta_D\left(\bm{x}-\bm{q}-D\bm{s}(\bm{q})\right)=\int \frac{d^3 q}{(2\pi)^3} e^{-i\bm{k}\cdot\left(D\bm{s}(\bm{q})+\bm{q}\right)}\ .
\ee

We can write the displacement field as a sum of a short-scale, $\bm{s}_s$, and a large-scale piece, $\bm{s}_\Lambda$. To lowest order in $k_{\rm long}$, the large-scale displacement can be written as (in Lagrangian space):
\be\label{largeS}
\bm{s}_{\Lambda}(\bm{q},\eta)\approx -\bm{q}\frac{1}{3 D(\eta)}\delta_{\Lambda,L}(\eta)\ ,
\ee
which corresponds to a $\delta_{\Lambda,L}$ being constant in space. Thus, in our calculation we only take into account the change in the local $\Omega_{\rm matter}$ induced by the presence of $\delta_{\Lambda,L}$. 

Then we can expand $\delta_Z$ to third order in $\bm{s}$ (i.e. third order in $\delta_L$), and use the split $\bm{s}=\bm{s}_s+\bm{s}_\Lambda$ together with eq.~(\ref{largeS}). After some algebra we obtain (in configuration space)
\be\label{deltaZmaster}
\delta_Z&=&\delta_{s,L}\left(
1+\frac{4}{3}\delta_{\Lambda,L}+\frac{10}{9}\delta_{\Lambda,L}^2
\right)\\\nonumber
&&+D\left(\bm{s}_\Lambda\cdot\nabla_{\bm{q}}\right)\delta_{s,L}-\frac{5}{3}D\delta_{\Lambda,L}\left(\bm{s}_\Lambda\cdot\nabla_{\bm{q}}\right)\delta_{s,L}
+\frac{D^2}{2}s_{\Lambda}^i s_{\Lambda}^j\partial_{q^i}\partial_{q^j}\delta_{s,L}\ .
\ee
Now let us show that the second line gives no contribution to the power spectrum $P_Z$ calculated to second order in $P_L$. When calculating $P_Z$, we use that the linear short-scale and large-scale modes are independent ($\la \delta_{s,L}\delta_{\Lambda,L}\ra$=0). Thus, we can see that the second term on the second line gives a contribution to $P_Z$ proportional to
$$
\la \delta_{\Lambda,L}s_{\Lambda}^i\ra
\la \delta_{s,L}\partial_{q^i}\delta_{s,L}\ra=0\ ,
$$
which vanishes because of the vanishing first bracket (by isotropy). 

Analogously, the first and third term on the second line of eq.~(\ref{deltaZmaster}) give equal but opposite in sign contributions to the power spectrum, which therefore cancel out as well. Thus, we can see that at this order $P_Z$ depends only on $\delta_{\Lambda,L}$ but not on the large-scale displacement field. This must indeed be the case since as we already argued equal-time statistics in the ZA must be independent of the large-scale bulk flows.

Thus, for the purposes of equal time statistics, we can simply write
\be\label{deltaZT}
\delta_Z=\delta_{s,L}\left(
1+\frac{4}{3}\delta_{\Lambda,L}+\frac{10}{9}\delta_{\Lambda,L}^2
\right)\ ,\ee
plus irrelevant bulk flow terms. The power spectrum is then given by
\be\label{PZPL}
P_Z(k,\eta)=P_L(k,\eta)\left(1+4 \delta^2(<\Lambda,\eta)\right)\ .
\ee
Expressing $P_L$ from the above equation and plugging it in eq.~(\ref{Pnlshort}) we recover eq.~(\ref{PnlshortZold}), which was the goal of this section.
Note that by comparing eq.~(\ref{deltaZT}) and eq.~(\ref{deltaSNL}) we obtain eq.~(\ref{sNLZ}), which is our model for the short-scale non-linear overdensity.

\acknowledgments ST would like to thank Daniel Eisenstein for many helpful conversations. The work of MZ is supported by NSF grants PHY-0855425, AST-0506556 \& AST-0907969, and by the David \& Lucile Packard and the John D. \& Catherine T. MacArthur Foundations.

\bibliography{mildly_NL_v1}

\end{document}